\documentclass{jfm}

\usepackage{graphicx}
\usepackage{natbib}

\usepackage{amsmath,amssymb,amsfonts}
\usepackage{color}
\usepackage{varioref}
\usepackage{wrapfig}
\usepackage[FIGTOPCAP, hang, raggedright, nooneline]{subfigure}
 \usepackage{subfigmat}
\usepackage{dcolumn}
\usepackage{bm}
\usepackage{multirow}
\usepackage{rotating}
\usepackage{url}
\usepackage{cancel}

\usepackage{tikz}

\usepackage[T1]{fontenc}

\ifCUPmtlplainloaded \else
  \checkfont{eurm10}
  \iffontfound
    \IfFileExists{upmath.sty}
      {\typeout{^^JFound AMS Euler Roman fonts on the system,
                   using the 'upmath' package.^^J}%
       \usepackage{upmath}}
      {\typeout{^^JFound AMS Euler Roman fonts on the system, but you
                   dont seem to have the}%
       \typeout{'upmath' package installed. JFM.cls can take advantage
                 of these fonts,^^Jif you use 'upmath' package.^^J}%
      }
  \else
  \fi
\fi

\ifCUPmtlplainloaded \else
  \checkfont{msam10}
  \iffontfound
    \IfFileExists{amssymb.sty}
      {\typeout{^^JFound AMS Symbol fonts on the system, using the
                'amssymb' package.^^J}%
       \usepackage{amssymb}%

      }{}
  \fi
\fi

\ifCUPmtlplainloaded \else
  \IfFileExists{amsbsy.sty}
    {\typeout{^^JFound the 'amsbsy' package on the system, using it.^^J}%
     \usepackage{amsbsy}}
    {}
\fi

\newsavebox{\astrutbox}
\sbox{\astrutbox}{\rule[-5pt]{0pt}{20pt}}

\makeatletter

\newcommand{\volumedash}{%
  \makebox[0pt][l]{%
    \ooalign{\hfil\hphantom{$\m@th V$}\hfil\cr\kern0.08em--\hfil\cr}%
  }%
}
\makeatother

\title[General governing equations for fluid discontinuities]{Capturing the kinematics and dynamics of fluid fronts}

\author[J. J. Thalakkottor and K. Mohseni]%
{J\ls O\ls S\ls E\ls P\ls H\ns J.\ns T\ls H\ls A\ls L\ls A\ls K\ls K\ls O\ls T\ls T\ls O\ls R$^1$\thanks{Email address for correspondence: joseph.johnthalakkottor@sdsmt.edu}%
  \ns
\and K.\ns M\ls O\ls H\ls S\ls E\ls N\ls I$^{1,2}$}

\affiliation{$^1$Department of Mechanical and Aerospace Engineering, University of Florida,\\ Gainesville, Florida 32611, USA\\[\affilskip]
$^2$Department of Electrical and Computer Engineering, University of Florida,\\ Gainesville, Florida 32611, USA}

\pubyear{2010}
\volume{650}
\pagerange{119--126}

\begin{document}

\maketitle

\begin{abstract} 
Gibbs was the first person to represent a phase interface by a dividing surface. He defined the dividing surface as a mathematical surface that has its own material properties and internal dynamics. In this paper, an alternative derivation to this mathematical surface is provided that generalizes the concept of dividing surface to fluid fronts beyond that of just a phase or material interface. Other fluid fronts being a vortex sheet, shock front, moving contact line, and gravity wavefront, to name a few. Here, this extended definition of dividing surface is referred to as the {\it extended dividing hypersurface} (EDH), as it is not just applicable to a surface front but also to a line and a point front. This hypersurface is a continuum approximation of a diffused region with fluid properties and flow parameters varying sharply but continuously across it. 
This paper shows that the properties and equations describing an EDH can be derived from the equations describing the diffused region by integrating it in the directions normal to the hypersurface. This is equivalent to collapsing the diffused region in the normal direction. Hence, ensuring that the EDH is both kinematically and dynamically equivalent to that of the diffused region. 
%
Various canonical problems are examined to demonstrate the ability of the EDH to accurately represent different types of fluid and flow fronts, including static and dynamic interfaces, shock fronts, and vortex sheets. These examples emphasize the EDH's capability to represent various functionalities within a front, the relationship between the flux of quantities and hypersurface quantities, and the importance of considering the mass of front and associated dynamics.

\end{abstract}


\section{Introduction}
A  {\it front} is often accompanied by rapid changes in scales, multiphysics, geometrical complexities, and intriguing chemical phenomena, making it an ideal benchmark to expand our knowledge beyond the confines of the continuum field.
A front refers to a boundary separating two or more sets of homogeneous quantities in a continuum field. Across the front, one or more of these field quantities are usually discontinuous. While some of these quantities pertain to the material properties of the media, others are associated with its kinematics and dynamics. Examples of commonly occurring fronts in fluid mechanics, across which fluid properties and/or flow parameters are discontinuous, are listed in table \ref{tab:Discontinuities}. 
These fronts can be categorized into two types. The first type is referred to as the physical front, which exhibits sharp gradients in fluid and flow parameters at length scales that are below the experimental or numerical observable limit \citep{Mohseni:10a}. Moreover, the equation of state and thermodynamic properties of the physical front differ from those of the surrounding homogeneous media \citep{AmsdenAA:71a, AbgrallR:91a}. Examples of physical fronts include material or phase interface, gravity wave front \citep{SuleC:93a,SutherlandBR:10a, NappoCJ:13a}, and shock front \cite{ShapiroAH:53a}.

The second type is referred to as the apparent front, characterized by gradients that exist at length scales above the observable limit but can be approximated as a discontinuity for analytical and numerical simplicity. In this case, the equation of state and the thermodynamic properties within the front are the same as the surrounding homogeneous media. Examples of apparent fronts include a vortex sheet and an entrainment sheet, which are commonly used to describe a boundary layer or mixing layer.
Among these examples of a front is a well-known discontinuity corresponding to the jump in pressure across a curved fluid-fluid interface. This jump in pressure is attributed to interfacial tension, which is the thermodynamic property of an interface in an equilibrium system. This then begs the following questions: (1) whether the fluid interface possesses other material properties? (2) does it have any internal dynamics if the interface is in a non-equilibrium system?, And (3) if the material and phase interface can have their own properties and internal dynamics, could other fluid and flow fronts also possess their own distinct material properties and internal dynamics? The fact that a fluid interface could have material properties analogous to that of a bulk fluid and its own internal dynamics was first recognized by \cite{GibbsJB:28a}.
\begin{table}\label{tab:Discontinuities}
\begin{center}
\def~{\hphantom{0}}
\begin{tabular}{l*{7}{c}r} 
  & ~Types of fronts / Discontinuity~ & ~$\rho$ & ~$p$~ & ~$T$~ & ~$u_s$~ & ~$u_n$~ & ~$\tau_{sn}$~  \\[10pt]
& {\bf Physical front }                               & ~ ~ & ~ ~ & ~  ~ & ~  ~ & ~  ~ & ~  ~  \\
& Hydrophilic/Miscible interface       		  & ~D~ & ~D~ & ~ - ~ & ~ - ~ & ~ - ~ & ~ - ~  \\
& Hydrophobic/Immiscible boundary    	  & ~D~ & ~D~ & ~D~ & ~ D ~ & ~ - ~ & ~ - ~  \\
& Moving contact line			    	  & ~D~ & ~D~ & ~D~ & ~ D ~ & ~ - ~ & ~ D ~  \\
& Interface with surface tension gradient     & ~ D ~ & ~ D ~ & ~ - ~ & ~ - ~ & ~ - ~ & ~ D ~  \\ 
& Contact discontinuity                	          & ~ D ~ & ~ - ~ & ~ D ~ & ~ D ~ & ~ - ~ & ~ - ~  \\
& Gravity wave front               	                  & ~ D ~ & ~ - ~ & ~ - ~ & ~ - ~ & ~ D ~ & ~ - ~  \\
& Shock front                                	          & ~ D ~ & ~ D ~ & ~ D ~ & ~ - ~ & ~ D ~ & ~ - ~  \\ \\
& {\bf Apparent (pseudo) front }                               & ~ ~ & ~ ~ & ~  ~ & ~  ~ & ~  ~ & ~  ~  \\

& Vortex sheet                                	          & ~ D ~ & ~ - ~ & ~ - ~ & ~ D ~ & ~ - ~ & ~ - ~  \\
& Entrainment sheet$^1$                             & ~ D ~ & ~ - ~ & ~ - ~ & ~ - ~ & ~ D ~ & ~ - ~  \\
  \end{tabular}
  \caption{\label{tab:Discontinuities} List of fronts and corresponding field quantities that are discontinuous across it. The fluid properties or flow parameters that are discontinuous are labelled as `D'. Here, $\rho$ is the density, $p$ pressure, $T$ temperature, $u_s$ tangential component of velocity, $u_n$ normal component of velocity, and $\tau_{sn}$ shear stress.}
\end{center}
\end{table}

Gibbs introduced the concept of a dividing surface to represent a fluid interface, more specifically, a phase interface.  The dividing surface is a two-dimensional mathematical surface with zero thickness, which has its own properties and internal dynamics \citep{GibbsJB:28a,ScrivenLE:60a,SlatteryJC:64a}. Gibbs proposed a phenomenological description of the thermodynamic relations for the dividing surface, which represents a phase interface in a body at rest or equilibrium.  After nearly three decades, \cite{ScrivenLE:60a} extended this concept of dividing surface to non-equilibrium systems by deriving a model that described the internal dynamics of the fluid within the dividing surface \citep{ScrivenLE:60a,ArisR:62a}. Though Scriven accounted for the coupling of the dividing surface with its surrounding bulk media, the mass transfer between the dividing surface and the bulk fluid surrounding it was ignored. The effect of mass transfer was later included in the governing equations by \cite{SlatteryJC:64a}. This was done by expanding Gibbs' definition of homogenous media to one where the constitutive equations apply uniformly. In these three models, the thermodynamic relations \citep{GibbsJB:28a}, the transport equations, and the conservation of mass, momentum, and energy of the dividing surface \citep{ScrivenLE:60a,SlatteryJC:64a} were all independently {\it defined} as a two-dimensional analogue to the corresponding three-dimensional bulk equations. The last question posed in the previous paragraph still remains unanswered, that is, if other fluid fronts can also be described by a dividing surface concept.

The present paper addresses this question by delving into the characteristics of fluid fronts. It is asserted that the actual physical front is a diffused region with three dimensions and a finite thickness. Within this region, fluid and flow parameters exhibit sharp but continuous variations across its width. It is important to note that the conventional representation of a front as a hypersurface in continuum theory is a limiting case of this diffused region. As a result, the authors propose a systematic derivation of the dividing surface from the 3D bulk conservation equations that accurately describe this diffused region. This generalized dividing surface is referred to as the extended dividing hypersurface (EDH). The EDH equations are derived by collapsing the dimension across the width of the diffused region, mathematically achieved through integration along its width. This mathematical treatment ensures that the EDH is kinematically and dynamically equivalent to the diffused region and represents the real physical front in its entirety.

To validate the EDH model and its generalization, the authors conduct a comprehensive analysis of canonical problems that involve fluid fronts. These problems are (1) stationary fluid with varying miscibility, (2) stratified flow through a converging-diverging section, (3) the shock tube problem, (4) the vortex entrainment sheet, and (5) unsteady bubble dynamics.

The selected problems in this study serve four main purposes. Firstly, they demonstrate that the extended dividing hypersurface (EDH) is capable of accurately capturing the dynamics of different types of fluid and flow fronts, not limited to phase or material interfaces. Secondly, they highlight that the EDH can effectively represent various functionalities within a front, going beyond the commonly described monotonicity distribution in literature, resulting in new classes of fronts. Thirdly, they illustrate the relationship between the flux of $m$-dimensional quantities and the $m$-1 dimensional quantities (referred to as hypersurface quantities), emphasizing how this coupling can lead to hypersurface dilatation even in incompressible hypersurface flows, which is counterintuitive to deductions made from continuity equations for bulk fluids. Moreover, the study emphasizes the importance of acknowledging the mass of the front and capturing the associated dynamics. 

The paper is outlined as follows: section 2 defines a diffused region, front, and hypersurface. Here, a brief description is provided of how the collapse of dimension is achieved mathematically, along with an overview of the methodology used to obtain the governing equations of an EDH. Section 3 goes through the derivation of equations pertaining to EDH. In section 4, the details of numerical simulations used for validation of the EDH model are presented. Finally, the results are presented and discussed in section 5.

\section{Overview of methodology and general definitions}
Prior to deriving the governing equations for the EDH, it is helpful to: 1) introduce key definitions and nomenclature utilized in this study, 2) provide an overview of the concept of spatial dimension collapse, and 3) outline the adopted methodology for deriving the governing equations for the EDH.

\subsection{Defining a diffused region, the hypersurface, and the extended dividing hypersurface}
As previously stated, a front is referred to as a fluid feature across which one or more fluid or flow parameters is {\it considered} to be discontinuous. This can be mathematically and computationally represented as a diffused region with finite volume or a hypersurface with zero thickness.

\subsubsection{Diffused region}
In this paper, a {\it diffused region} is defined as a region with finite thickness in $m$-dimensional space, where one or more field quantities, such as fluid properties or flow parameters, exhibit sharp but continuous variations. The diffused region serves as a more realistic representation of a front, as true mathematical discontinuities seldom exist in the physical world.

\begin{figure}
\begin{minipage}{1.0\linewidth}
(a) $m$ = 3
\end{minipage}
\begin{minipage}{1.0\linewidth}
\centerline{
 \includegraphics[width=1.0\textwidth]{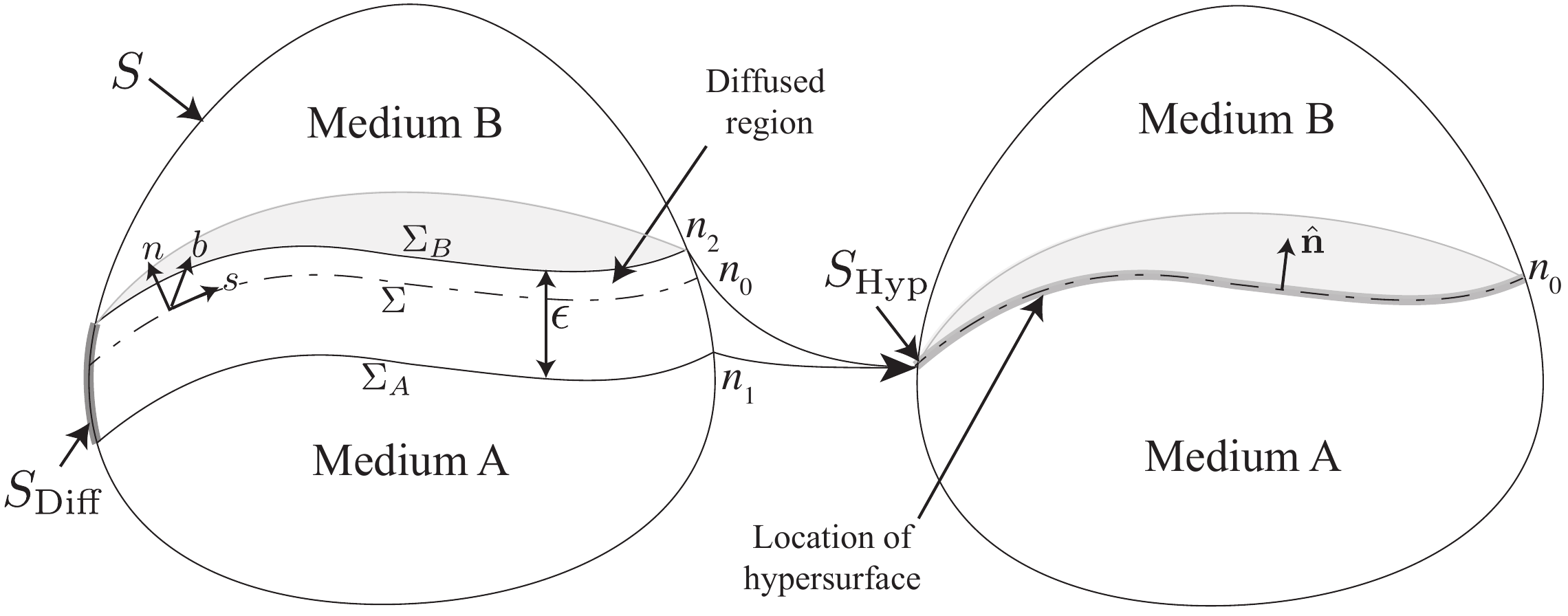}
 }
 \end{minipage}\\
\\
\\
 \begin{minipage}{1.0\linewidth}
(b) $m$ = 2
\end{minipage}
\begin{minipage}{1.0\linewidth}
\centerline{
 \includegraphics[width=1.0\textwidth]{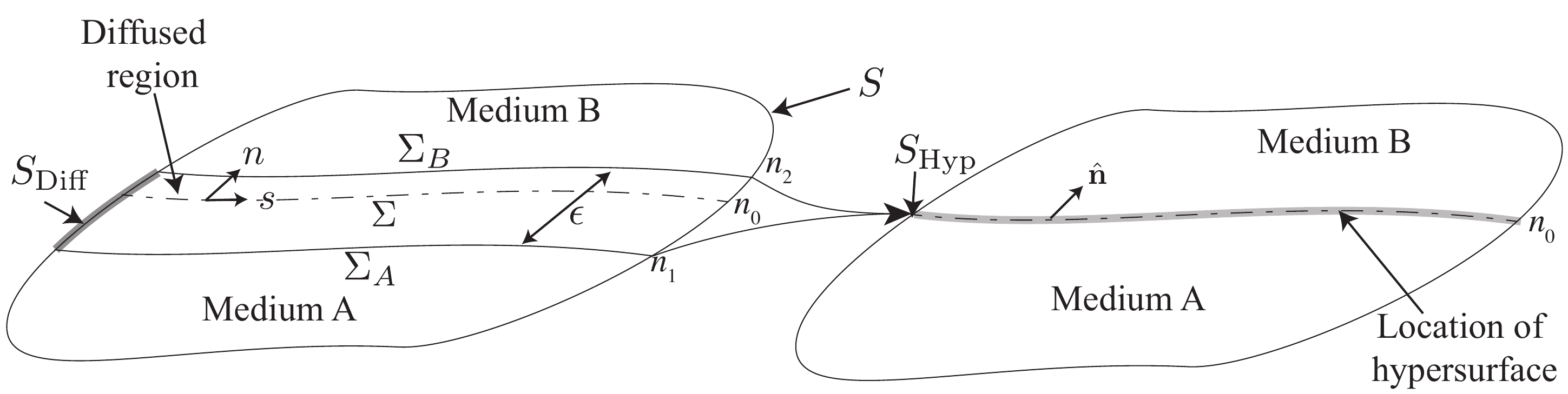}
 }
 \end{minipage}\\
\\
\\
  \begin{minipage}{1.0\linewidth}
 (c) $m$ = 1
\end{minipage}
\begin{minipage}{1.0\linewidth}
\centerline{
 \includegraphics[width=0.7\textwidth]{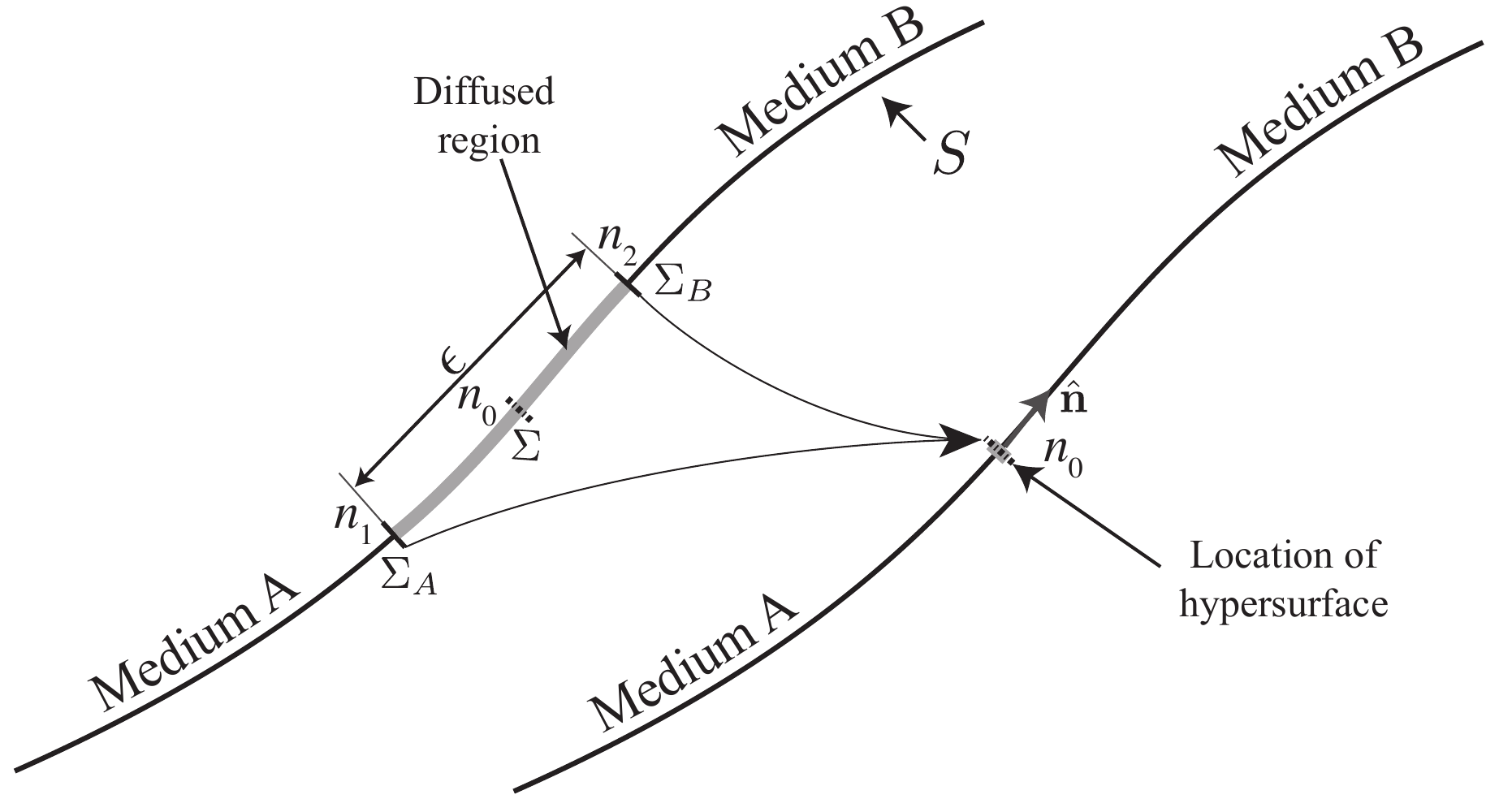}
 }
 \end{minipage}\\
  \caption{Schematic of an arbitrary control (a) volume, (b) surface, and (c) line, depicting two homogeneous media separated by a diffused region or a hypersurface. The boundary of the domain encompassing the homogenous media and the embedded diffused region or the hypersurface, is denoted by `$S$'. The boundary `S' is further divided into `$S_{\text{Bulk}}$' and `$S_{\text{Diff}}$', corresponding to the sections of the boundary covering the homogeneous media and the diffused region, respectively.   $\Sigma$ is used to represent the EDH. The boundary of the hypersurface is denoted by $S_{\text{Hyp}}$.  The bounds of the diffused region, separating it from the homogeneous media $A$ and $B$ are named as $\Sigma_{\text{A}}$ and $\Sigma_{\text{B}}$, respectively. The width of the diffused region is denoted by $\epsilon$. The location of the $\Sigma_{\text{A}}$, $\Sigma_{\text{B}}$, and $\Sigma$ are given as $n_2$, $n_1$, and $n_0$, respectively.
}
\label{fig:Schematic-Arbitrary_Diff_Intf}
\end{figure}

The diffused region is depicted in Figure \ref{fig:Schematic-Arbitrary_Diff_Intf}. The boundary of the domain containing the homogeneous media and the embedded diffused region or hypersurface is represented by `$S$'. `$S$' is further divided into `$S_{\text{Bulk}}$' and `$S_{\text{Diff}}$', corresponding to the sections of the boundary covering the homogeneous media and the diffused region, respectively. The boundaries of the diffused region that separate it from the homogeneous media $A$ and $B$ are identified as $\Sigma_{\text{A}}$ and $\Sigma_{\text{B}}$, respectively. The locations of $\Sigma_{\text{B}}$ and $\Sigma_{\text{A}}$ are indicated as $n_2$ and $n_1$, respectively. The width of the diffused region is denoted by $\epsilon$. It must be noted that these are not sharp, distinct boundaries of the diffused region but rather an apparent boundary that are set based on criteria of a fluid or flow parameter defined apriori. This is analogous to how velocity boundary layer and its bounds are defined as the location where fluid velocity is $99 \%$ of the free stream velocity.

%

\subsubsection{Hypersurface}

 In mathematics, a hypersurface is a manifold of dimension $m-1$ that is embedded in a $m$-dimensional ambient space. In other words, referring to figure \ref{fig:Schematic-Arbitrary_Diff_Intf}, the hypersurface in a $3$-D space is a surface, in a $2$-D space is a curve, and in a $1$-D space is a point. Hence, in contrast to the diffused region, a hypersurface has zero thickness. Representing a front as a hypersurface, results in a true mathematical discontinuity across it. This approach is commonly employed for representing fronts due to its simplicity.

\subsubsection{Extended Dividing Hypersurface (EDH)}
An extended dividing hypersurface is defined as a type of hypersurface that accurately represents and encompasses the cumulative (integrated in the direction normal to the front) kinematics and dynamics of a front as a diffused region. The EDH distinguishes itself from a diffuse region by its zero thickness in the front normal direction. Across the EDH, the field quantities undergo discontinuous changes rather than continuous variations. In comparison to a Gibbs' dividing surface, the EDH separates not only two homogeneous media but also the same media at different dynamic states. Finally, the EDH serves as an extension to the Gibbs' dividing surface, which was initially limited to representing material or phase interfaces. The EDH, in contrast, offers the flexibility to represent any fluid front.  Examples of such fronts include vortex sheets, shock fronts, moving contact lines, and triple points. Hence, the term "hypersurface" is preferred over "surface" as it encompasses not only geometric surfaces but also fronts that exist as curves, lines, and points.

In figure \ref{fig:Schematic-Arbitrary_Diff_Intf}, $\Sigma$ is used to represent the EDH, with its location given by $n_0$. The boundary of the hypersurface is denoted by $S_{\text{Hyp}}$.  The normal vector $\hat{\bf n}$ to the hypersurface is defined as the direction along which sharp variation occurs within the diffused region. Based on it, a right-handed orthogonal coordinate system is defined $(s, n, b)$ on the hypersurface, with $\hat{\bf s}$ and $\hat{\bf b}$ as the basis vectors, locally tangent to the hypersurface

\subsection{Overview of collapsing a spatial dimension mathematically}
 \begin{figure}
\begin{minipage}{1.0\linewidth}
\centerline{
 \includegraphics[width=0.75\textwidth]{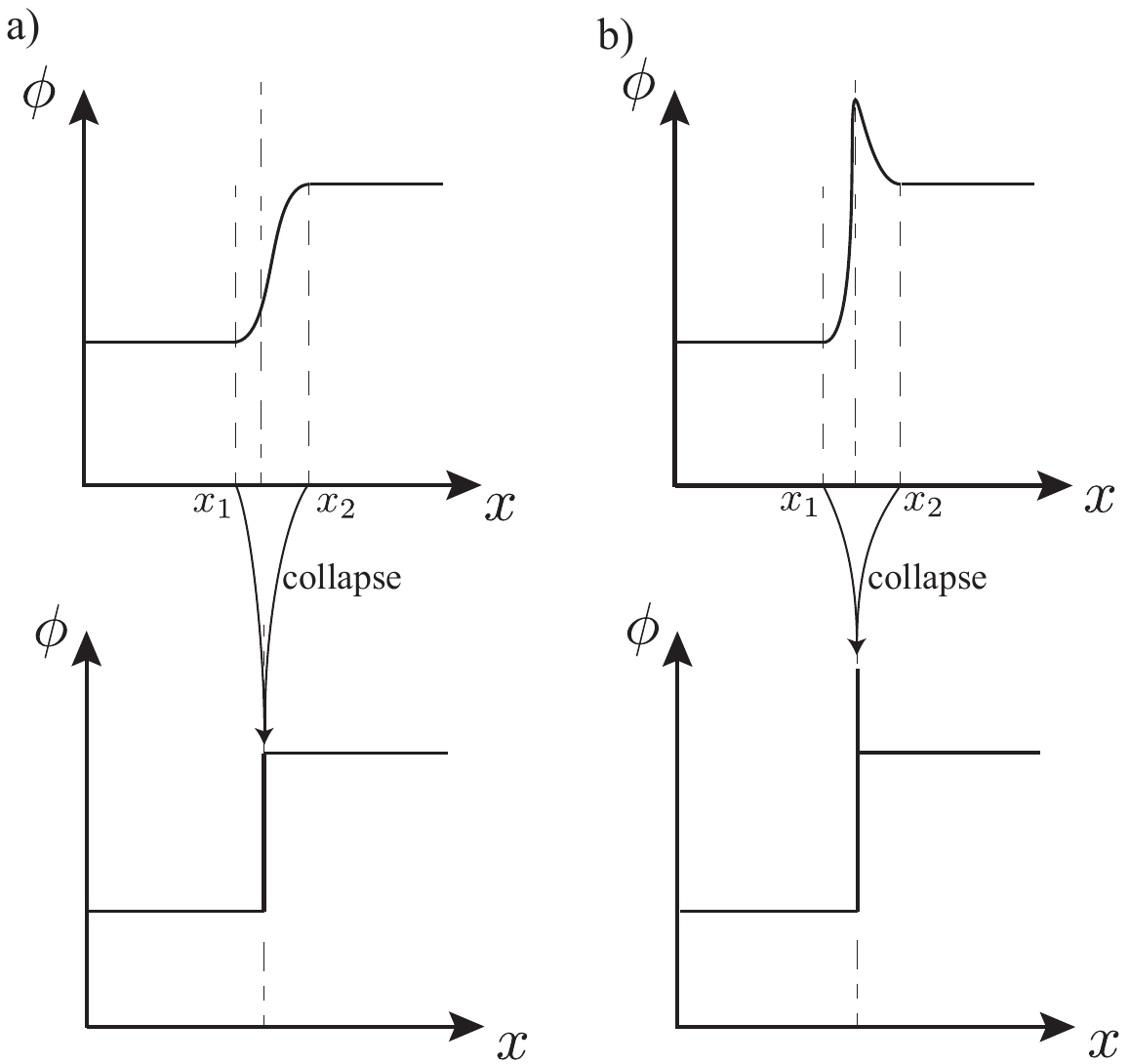}
 }
 \end{minipage}
  \caption{The schematic demonstrates two cases with the same states in the bulk homogenous media, in which the diffused region, where $\phi$ varies sharply but continuously, is collapsed to a mathematical hypersurface. Note that even though the states of bulk media are the same in both cases, the hypersurfaces are different. This collapse in dimension is performed by integrating $\phi$ across the diffused region. The lumped value of $\phi$ in this lower dimension is hence given as $\phi^{(m-1)}=\int_{x_1}^{x_2} \phi dx$. In the first case (a), $\phi$ is a monotonic function, whereas the (b) second case is not.}
  \label{fig:SchemCollapse}
\end{figure}

The method of collapsing a dimension can be elucidated using a schematic, as depicted in figure \ref{fig:SchemCollapse}. We examine two scenarios characterized by distinct distributions of an arbitrary parameter $\phi$ within the diffused region, while exhibiting identical jumps in $\phi$ across this region.  
Figure \ref{fig:SchemCollapse}  (a) displays a monotonic variation of $\phi$ between the two limits, as commonly observed in the literature.
In figure \ref{fig:SchemCollapse} (b),  $\phi$ is not monotonic. In this case, the average is not constrained to be within the two limiting values. The latter is also a physically plausible distribution, as shall be demonstrated for the case of partially miscible fluids (Section \ref{subsec_1}). It can be shown that the diffused region can be collapsed by integrating the distribution of $\phi$ across the width of the diffused region $\left(\int_{x_1}^{x_2} \phi dx\right)$, while simultaneously preserving the total value of $\phi$. In other words, the dimension can be collapsed by integrating field quantities in the direction of the collapsing dimension or in the direction normal to the hypersurface. {\it The knowledge of the integrated or lumped value of $\phi$ and its functionality within the diffused region helps uniquely identify a front.}

\subsection{Overview of deriving hypersurface equations}

\begin{figure}
\begin{minipage}{1.0\linewidth}
\centerline{
 \includegraphics[width=1.0\textwidth]{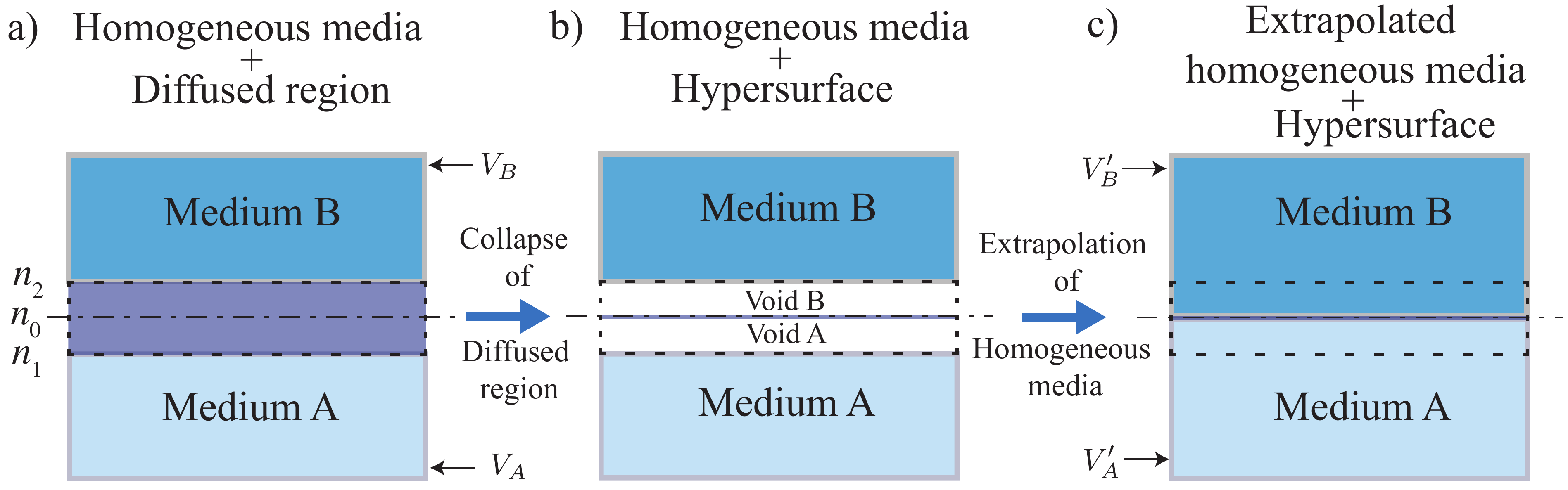}
 }
 \end{minipage}
   \caption{Schematic depicts the three steps required to derive the equations describing an {\it extended dividing hypersurface} from the equations for a diffused region.
 The figure (a) shows the diffused region with a finite thickness. The figure (b) shows the extended dividing hypersurface with zero thickness, achieved by collapsing the diffused region and the corresponding void created as a result of it. The figure (c) presents the EDH embedded in a homogenous media with the void now being occupied by the two homogeneous media.}
\label{fig:Schematic-Rectangle_Diff_Intf1}
\end{figure}

The derivation of governing equations for an extended dividing hypersurface (EDH) is based on the fundamental objective of capturing the cumulative kinematics and dynamics of a front as a diffused region. To accomplish this, the equations governing the EDH are derived from those of the diffused region. This process can be divided into three distinct steps, which are as follows:
\begin{enumerate}
\item {\it Describe the diffused region} (Section \ref{Sec:Mass_Conserv_Diff}). As previously mentioned, we view the diffused region of finite thickness, in $m$-dimensional space, as being the closest physical representation of the front,  figure \ref{fig:Schematic-Rectangle_Diff_Intf1} (a). Hence, the first step is to present all the necessary equations needed to completely and uniquely define a diffused region. This set of equations includes governing equations, boundary, and initial conditions.

\item {\it Derive equations for an actual extended dividing hypersurface, by collapsing the diffused region} (Section \ref{Sec:SharpHypSurf_Act}). The diffused region is collapsed in dimensions normal to the EDH by integrating the equations describing it, along these directions, as seen in figure \ref{fig:Schematic-Rectangle_Diff_Intf1} (b). This results in a set of equations describing an EDH. Therefore, in the EDH, the various flow and fluid quantities associated with the diffused region are now treated as quantities lumped in the normal direction. In this paper this collapsed diffused region is referred to as the {\it actual} extended dividing hypersurface. This is because often, we are not concerned with just modeling the front as an isolated entity, but as an entity embedded in a homogeneous media, for which one last step needs to be done.

\item {\it Derive equations for an effective extended dividing hypersurface} (Section \ref{Sec:SharpHypSurf_Eff}). As a consequence of the collapse, a void is created in the space previously occupied by the diffused region. This void is considered to be occupied by the adjacent homogeneous media of $m$-dimensional space (figure \ref{fig:Schematic-Rectangle_Diff_Intf1} (c)). Since we intend to model a homogeneous system with an embedded hypersurface, this added homogeneous matter and its dynamics needs to be accounted and adjusted for. This is done by subtracting the equation describing this added matter from the equations for the {\it actual} EDH. This gives us the equations for what here will be referred to as the {\it effective} extended dividing hypersurface. This ensures that the system of homogeneous media with an embedded hypersurface (figure \ref{fig:Schematic-Rectangle_Diff_Intf1} (c)) is both kinematically and dynamically equivalent to that of the homogeneous media with the diffused region (figure \ref{fig:Schematic-Rectangle_Diff_Intf1} (a)). 
\end{enumerate}

With this basic framework in mind, the equations for an EDH can be derived. In order to completely and properly describe a fluid system, a set of governing equations, boundary, and initial conditions are required. The governing equations comprise of the conservation of mass, momentum, and energy, and the equation of state. In order to ensure there is no loss of generality, the equations are stated for a region in an arbitrary $m$-dimensional space. As a result, this derivation is not just limited to finding the equations of a $2$-D dividing surface from a $3$-D region, but is also applicable to finding equations for a $1$-D dividing line or $0$-D dividing point.
We detail out the steps of this derivation in the following section, using mass conservation as an example. Similar steps can then be used to derive the remaining governing equations, boundary, and initial conditions.

\section{Deriving equations for hypersurfaces/Mathematical formulations}

\subsection{Deriving mass conservation for an extended dividing hypersurface}
As previously mentioned, the derivation of governing equations for an {\it extended dividing hypersurface} can be divided into three steps. The first of the three steps is describing the diffused region. %

\subsubsection{Describing the mass conservation for a diffused region \label{Sec:Mass_Conserv_Diff}} 

In the case of a fluid media containing a diffused region, a single set of governing equation is used to describe the homogeneous media and the diffused region in $m$-D space. This is realized by allowing the material properties to vary from one value of the homogeneous media to the other, through the diffused region. With this in mind, the mass conservation equation for a diffused region is presented below in the conservative integral form.

%
\begin{equation}
\frac{\text{d}}{\text{d}t} (M) = \Pi_{\text{mass}}.
\end{equation}
Here, $M$ is the total mass in the material region and $\Pi_{\text{mass}}$ is the net source of mass when applicable. A relation in terms of field quantities is obtained, by writing the total mass of the region in terms of the local density, $\rho$, and the source density term as  $\pi_{\text{mass}}^m$. 
\begin{equation}
\frac{\text{d}}{\text{d}t} \left(\int_{m} \rho^m dx_m \right) = \int_{m} \pi_{\text{mass}}^m dx_m .
\end{equation}
Here, considering an orthogonal coordinate system, $\int_{m} \equiv \int_1\int_2\int_3....\int_m$ and $dx_m \equiv dx_1dx_2dx_3....dx_m$. The  superscripts $(\cdot)^m$ and subscripts $(\cdot)_m$, denotes fluid quantities and operators in $m$-dimensional space, respectively.
Using the Reynolds' transport theorem (Leibniz's integral rule), the above equation becomes:
\begin{align}
\int_{m} \left\{\partial_t \left( \rho^m \right) + \bm\nabla^m \bm\cdot \left(\rho\bf u \right)^m = \pi_{\text{mass}}^m \right\} dx_m .
\label{eq:MassConsv_RefCoord}
\end{align}
Here, $\partial_t = \partial/\partial t$ is the derivative with respect to time. This is the most fundamental representation of the principle of mass conservation. 
With the mass conservation for a diffused region presented, we next move on to derive the mass conservation for an {\it actual} extended dividing hypersurface.

\subsubsection{\it Derive mass conservation for an actual extended dividing hypersurface, by collapsing the diffused region \label{Sec:SharpHypSurf_Act}} 
Our objective is to derive conservation equations for an EDH which is kinematically and dynamically equivalent to that  for a diffused region. The mass conservation equation for the EDH is obtained by collapsing the  dimensions normal to the hypersurface. This is mathematically performed by integrating the conservation equation for a diffused region in that direction. 

The mass conservation for a hypersurface is evaluated by first splitting the gradient operator ($\nabla^m$) in the mass conservation into tangential and normal components.
This is done, because after the collapse, they will end up corresponding to terms representing the  flux of hypersurface quantities and bulk quantities.  For simplicity, the conservation equation in the reference coordinate system (equation \ref{eq:MassConsv_RefCoord}) is used. 
Hence, on decomposing equation \eqref{eq:MassConsv_RefCoord} we obtain:
\begin{align}
&\int_{m} \left( \partial_t \left(\rho^m \right) + \left[ \left({\bf I}^m - \hat{\bf n}\otimes\hat{\bf n}\right) \bm\cdot \nabla^m \right]  \bm \cdot (\rho{\bf u})^m +   \left[ \left(\hat{\bf{}n}\otimes\hat{\bf{}n}\right) \bm\cdot \nabla^m \right]  \bm \cdot (\rho{\bf u})^m  =  \pi_{\text{mass}}^m  \right) dx_m.
\end{align}
Here, $\left({\bf I}^m - \hat{\bf{}n}\otimes\hat{\bf{}n}\right) $ and $ \left(\hat{\bf{}n}\otimes\hat{\bf{}n}\right)$ denote the projection tensors in the tangent and normal directions to the hypersurface. 

\begin{figure}
\begin{minipage}{1.0\linewidth}
\centerline{
 \includegraphics[width=1.0\textwidth]{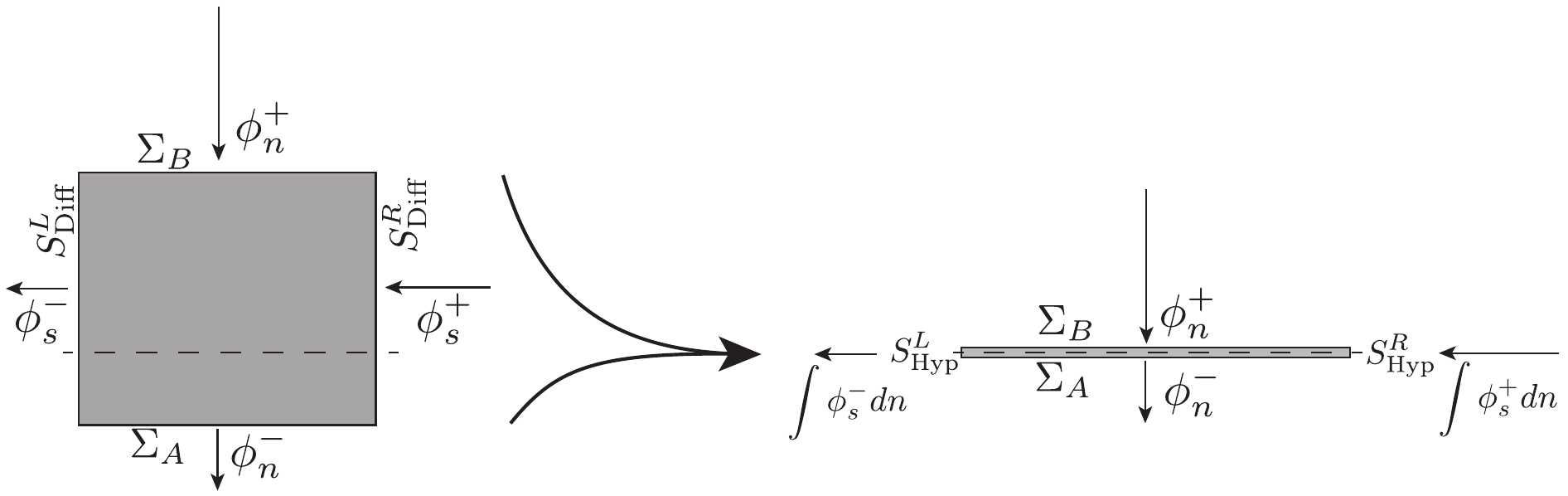}
 }
 \end{minipage}
   \caption{Schematic of a rectangular diffused region being collapsed in the normal direction to a hypersurface. }
\label{fig:Schematic-Collapse}
\end{figure}

The reason for this decomposition of the gradient operator can be further explained by using a rectangular diffused region as an example, as seen in figure \ref{fig:Schematic-Collapse}. The tangential component of the gradient operator is used to denote the mass flux through left and right boundaries of the diffused region ($S_{\text{Diff}}^L$ and $S_{\text{Diff}}^R$), while the normal component denotes the flux through the top and bottom boundaries ($\Sigma_A$ and $\Sigma_B$). 
On collapsing the diffused region by integrating across it, the flux in the tangential direction now corresponds to flux of lumped integral quantities $\left(\int \phi_s^{-}, \int \phi_s^{+}\right)$ through the boundaries of the hypersurface, $S_{\text{Hyp}}^L$ and $S_{\text{Hyp}}^R$. It will be later identified that the lumped quantities denote hypersurface quantities.
 As for the flux in the normal direction, it will result in the mass flux of homogeneous media into the EDH. Hence, the latter term results in the coupling between the flow in $m$-dimensional space with the flow in $(m-1)$-dimensional space. 

With the gradient operator decomposed, the integral in the direction normal to the hypersurface can be now separated:
\begin{align}
\int_{(m-1)}\left[  \int_{n_1}^{n_2} \left\{ \overbrace{\partial_t\left(\rho^m \right)}^{I}  + \overbrace{\nabla_{\hat{\bm s}}^{m-1} \bm \cdot (\rho{\bf u})^m}^{II} +   \overbrace{\nabla_{\hat{\bm n}}^m  \bm \cdot (\rho{\bf u})^m}^{III}  =   \overbrace{\pi_{\text{mass}}^m}^{IV} \right\} dn \right] dx_{(m-1)}.
\end{align}
Here, $\nabla_{\hat{\bm s}}^{m-1} = \left[ \left({\bf I}^m - \hat{\bf{}n}\otimes\hat{\bf{}n}\right)\bm\cdot \nabla^m \right] $ is the tangential projection of the gradient operator, $\nabla_{\hat{\bm n}}^m = \left[ \left(\hat{\bf{}n}\otimes\hat{\bf{}n}\right) \bm\cdot \nabla^m \right]$ is the normal projection of the gradient operator, and $n_1(s,b,t)$ and $n_2(s,b,t)$ are the location of the spatial bounds of the diffused region $\Sigma_{\text{A}}$ and $\Sigma_{\text{B}}$, respectively (figure \ref{fig:Schematic-Arbitrary_Diff_Intf}). The operator $\nabla_{\hat{\bm s}}^{m-1}$ represents a hypersurface gradient operator. The hypersurface gradient is the gradient operator in $(m-1)$-dimensional space. For the sake of clarity each term in the above equation is considered separately. For the first two terms, the integral is taken inside the temporal and spatial derivatives using Leibniz's rule: 
\begin{subequations}
\begin{align}\label{eq:Surf_Mass_1a}
 I \equiv \int_{n_1}^{n_2}  \partial_t\left(\rho^m \right) dn = \partial_t \left(\int_{n_1}^{n_2}  \rho^m dn \right)  - \left[
 \rho^m_{n_2} \frac{\partial {n_2}}{\partial t} -  \rho^m_{n_1} \frac{\partial {n_1}}{\partial t}\right]
\end{align}
The term $(\rho )^m_{n_2} \frac{\partial {n_2}}{\partial t} -  (\rho)^m_{n_1} \frac{\partial {n_1}}{\partial t}$ accounts for the flux due to the bounds of the diffused region varying with time. Next, we have%
\begin{align}\label{eq:Surf_Mass_1b}
II &\equiv  \int_{n_1}^{n_2} \left( \nabla_{\hat{\bm s}}^{m-1} \bm \cdot (\rho{\bf u})^m \right) dn   \notag \\ 
&=  \nabla_{\hat{\bm s}}^{m-1} \bm \cdot \left(\int_{n_1}^{n_2} (\rho{\bf u})^m dn\right) -  
 \left[ (\rho{\bf u})^m_{n_2} \bm\cdot \left(\nabla_{\hat{\bm s}}^{m-1} n_2 \right)  -   (\rho{\bf u})^m_{n_1} \bm\cdot \left(\nabla_{\hat{\bm s}}^{m-1} n_1 \right)\right]
\end{align}
 The term $  -(\rho{\bf u})^m_{n_2} \bm\cdot \left(\nabla_{\hat{\bm s}}^{m-1} n_2 \right)  +   (\rho{\bf u})^m_{n_1} \bm\cdot \left(\nabla_{\hat{\bm s}}^{m-1} n_1 \right)$ accounts for the flux due to the spatial variation in the bounds of the diffused region, along the hypersurface. In other words, it accounts for the changes in the width of the diffused region along the hypersurface. It must be mentioned that in the literature, it is usually assumed that the bounds are always parallel to the hypersurface (or the width of the diffused region is a constant) \citep{GibbsJB:28a,ScrivenLE:60a,SlatteryJC:64a} and moving with the same velocity as the hypersurface \citep{ScrivenLE:60a,SlatteryJC:64a}.
 
The third term can be exactly evaluated:
\begin{equation}
III \equiv \int_{n_1}^{n_2} \nabla_{\hat{\bm n}}^m  \bm \cdot (\rho{\bf u})^m dn =  \left[\!\left[ (\rho{\bf u})^m \bm\cdot \hat{\bf n} \right]\!\right]_{n_1}^{n_2}.
\end{equation}
Here, $\left[\!\left[ (\rho {\bf u})^{m} \right]\!\right]_{n_1}^{n_2} = \left(\rho {\bf u} \right)^m_{n_2} - \left(\rho {\bf u} \right)^m_{n_1}$ denotes the mass flux of the homogeneous media at the two bounds of the diffused region located at $n_1(s, b, t)$ and $n_2(s, b, t)$, where $(\rho{\bf u})^m_{n_1}=(\rho{\bf u})^m_{A,n_1}$ and $(\rho{\bf u})^m_{n_2}=(\rho{\bf u})^m_{B,n_2}$. This jump, represents a flow of information from the homogenous media in $m$-dimensional space to the EDH in $(m-1)$-dimensional space. Alternatively, this helps in viewing the EDH equations as a boundary condition to the homogeneous media in $m$-dimensional space. 
 The difference between the term $\frac{\partial {n}}{\partial t}$ in $I$ and the term $\bf{u}$ in $II$ is that the first corresponds to the velocity of the interface bounds while the other provides the velocity of the fluid at the location of interface bounds. When there is no mass flux through either of these locations, then the velocity of the interface bounds would match that of the velocity of fluid there.
 
As for the final term, it stays the same:
\begin{equation}
IV \equiv \int_{n_1}^{n_2}  \pi_{\text{mass}}^m dn.
\end{equation}
\end{subequations}

 Finally, the diffused region is collapsed in dimensions normal to the hypersurface by integrating each term in this direction. Collecting terms $I$, $II$, $III$ and $IV$ yields the mass conservation equation describing an {\it actual} hypersurface with zero thickness:   
\begin{align}\label{eq:Act_Surf_Mass_2}
& \int_{m-1}\left[ \partial_t \left( \rho^{(m-1)'} \right) \right.
 + \nabla_{\hat{\bm s}}^{m-1} \bm \cdot (\rho{\bf u})^{(m-1)'} - \pi_{\text{mass}}^{(m-1)'}    \\ \notag 
 &-\left[\!\left[ \rho^m \frac{\partial {n}}{\partial t}\right]\!\right]_{n_1}^{n_2}  -  \left[\!\left[ (\rho{\bf u})^m \bm\cdot \left(\nabla_{\hat{\bm s}}^{m-1} (n) \right) \right]\!\right]_{n_1}^{n_2}  +
 \left. \left[\!\left[ (\rho{\bf u})^m \bm\cdot \hat{\bf n} \right]\!\right]_{n_1}^{n_2} =  0 \right]dx_{m-1}.
\end{align}
%
Here, $ (\cdot)^{(m-1)'} = \int_{n_1}^{n_2}(\cdot)^mdn$ denotes {\it actual} hypersurface quantities. This equation provides the mass conservation for a hypersurface, but excludes any information regarding the void created by the collapse of the diffused region (referring back to figure \ref{fig:Schematic-Rectangle_Diff_Intf1}).  This void in the system is addressed next.

\subsubsection{Derive mass conservation for an effective extended dividing hypersurface \label{Sec:SharpHypSurf_Eff}} 
In this paper it is considered that the void is replaced by the homogeneous media adjacent to it. Therefore, an effective EDH adjusts for the effect of the dynamics created by the homogeneous media that ends up occupying this void. 
Before evaluating the effective EDH, we first present the equation of mass conservation for this added homogeneous fluid in lumped form:
\begin{align}\label{eq:Bulk_Mass_A1}
&  \int_{m-1}\left[ \partial_t \left( \rho^{(m-1)'}_A \right) \right.
+ \nabla_{\hat{\bm s}}^{m-1} \bm \cdot (\rho{\bf u})^{(m-1)'}_A - \pi^{(m-1)'}_A  \\ \notag
&-\left[\!\left[ \rho_A^m \frac{\partial {n}}{\partial t}\right]\!\right]_{n_1}^{n_0}  -  \left[\!\left[ (\rho{\bf u})_A^m \bm\cdot \left(\nabla_{\hat{\bm s}}^{m-1} (n) \right) \right]\!\right]_{n_1}^{n_0} +
 \left. \left[\!\left[ (\rho{\bf u})_A^m \bm\cdot \hat{\bf n} \right]\!\right]_{n_1}^{n_0} =  0 \right]dx_{m-1}.
\end{align}
%
Here, the subscripts $A$ denote the hypersurface quantities, equivalent to the newly added volume of homogeneous fluid A. 
The region occupied by it extends from $n_1$, the location of the bound of the diffused region next to medium A ($\Sigma_A$), to $n_0$ the location of the EDH. 

Similarly, for fluid B. 
\begin{align}\label{eq:Bulk_Mass_B1}
&  \int_{m-1}\left[ \partial_t \left( \rho^{(m-1)'}_B \right) \right.
+ \nabla_{\hat{\bm s}}^{m-1} \bm \cdot (\rho{\bf u})^{(m-1)'}_B - \pi^{(m-1)'}_B  \\ \notag
&-\left[\!\left[ \rho_B^m \frac{\partial {n}}{\partial t}\right]\!\right]_{n_0}^{n_2}  -  \left[\!\left[ (\rho{\bf u})_B^m \bm\cdot \left(\nabla_{\hat{\bm s}}^{m-1} (n) \right) \right]\!\right]_{n_0}^{n_2} +
 \left. \left[\!\left[ (\rho{\bf u})_B^m \bm\cdot \hat{\bf n} \right]\!\right]_{n_0}^{n_2} =  0 \right]dx_{m-1}.
\end{align}

This added material, and the dynamics as a result of it, needs to be adjusted for, in order for the homogeneous system with an embedded hypersurface to be identical to the system with a diffused region, which it is trying capture. Otherwise you would be double counting mass. This is taken care of by subtracting this additional contribution (equation \ref{eq:Bulk_Mass_A1} and \ref{eq:Bulk_Mass_B1}) from that of the {\it actual} EDH (equation \ref{eq:Act_Surf_Mass_2}). The resulting equation is called the mass conservation for an {\it effective} EDH or will be referred to here as just the mass conservation for an EDH:
%
\begin{align}\label{eq:Effect_Surf_Mass_2}
&{\partial_t \left( \rho^{(m-1)} \right)}
 + {\nabla_{\hat{\bm s}}^{m-1} \bm \cdot (\rho{\bf u})^{(m-1)}}  - {\pi_{\text{mass}}^{(m-1)}}  \\ \notag 
& -\left[\!\left[\rho^m  \frac{\partial {n}}{\partial t} \right]\!\right]_{n_0}- \left[\!\left[(\rho{\bf u})^m\right]\!\right]_{n_0} \bm\cdot \left(\nabla_{\hat{\bm s}}^{m-1} (n_0) \right)
 +{\left[\!\left[ (\rho{\bf u})^m \bm\cdot \hat{\bf n} \right]\!\right]}_{n_0} =   0 . 
\end{align}
Here, $\left[\!\left[ (\cdot)^{m} \right]\!\right]_{n_0} = \left(\cdot \right)^m_{B, n_0} - \left(\cdot \right)^m_{A, n_0}$ jump across EDH, the superscript $(m-1)$ corresponds to quantities associated with effective EDH. In the literature, the location at which the homogeneous quantities in the jump terms are evaluated has remained uncertain. Scriven suggested it is evaluated at the bounds of the diffused region, $\Sigma_A$ and $\Sigma_B$, while Slattery suggested it should be evaluated at the hypersurface, $\Sigma$. Here, we are able to analytically show that it is, in fact evaluated at the hypersurface, $\Sigma$.
In the above simplification, it is assumed that the mass flux is continuous across $\Sigma_A$ and $\Sigma_B$, that is, the mass flux on the homogeneous side of $\Sigma_A$ and $\Sigma_B$ is equal to the mass flux on the side of the diffused region.  Although this may seem obvious, it is, in fact, an assumption which is not always true. For example, when the hypersurface is separating a fluid and a solid media, there is a jump in density, $\left[\!\left[\rho \right]\!\right]$. Another example is when velocity slip occurs at the fluid-solid boundary \citep{Mohseni:13g,Mohseni:16b}, then there is a jump in the tangential component of velocity, $\left[\!\left[{ u} \right]\!\right]$.

Finally, observing that time derivative of $n_0$  (the location of the hypersurface) denotes the velocity of the hypersurface, ${\partial n_0}/{\partial t} = {\bf v}$, it allows us to rewrite the above equation as:
\begin{align}\label{eq:Effect_Surf_Mass_2}
&\overbrace{\partial_t \left( \rho^{(m-1)} \right)}^{\substack{\text{rate of change of} \\ \text{hypersurface mass}}}
 + \overbrace{\nabla_{\hat{\bm s}}^{m-1} \bm \cdot (\rho{\bf u})^{(m-1)}}^{\substack{\text{hypersurface mass} \\ \text{flux}}}  = \overbrace{\pi_{\text{mass}}^{(m-1)}}^{\substack{\text{source of} \\ \text{hypersurface mass}}}   \\ \notag 
&  +  \overbrace{\left[\!\left[(\rho{\bf u})^m\right]\!\right]_{n_0} \bm\cdot \left(\nabla_{\hat{\bm s}}^{m-1} (n_0) \right)}^{\substack{\text{flux of bulk mass due to} \\ \text{relative motion of hypersurface bounds}}} 
 - \overbrace{\left[\!\left[ (\rho ({\bf u} - {\bf v}))^m \bm\cdot \hat{\bf n} \right]\!\right]_{n_0}}^{\substack{\text{net bulk mass} \\ \text{entering hypersurface}}}. 
\end{align}
The first line in the above equation is analogous to the standard mass conservation represented in $(m-1)$-dimensional space. The second line on the other hand, accounts for the net mass flux of homogeneous fluid A and B in $m$-dimensional space, into the EDH.
Quantities associated with this effective hypersurface, are commonly referred to in the literature as surface quantities, when referring to 2D dividing surfaces \citep{SlatteryJC:64a}. This forms the basis of Gibbs definition for surface quantities which are often referred to as integral of excess quantities relative to the corresponding homogeneous quantity \citep{SlatteryJC:64a}. This is the final form of mass conservation equation that should be used while modeling an EDH embedded in a homogeneous fluid. 
Taking similar steps for momentum and energy equations, the conservation equations for effective hypersurface are obtained. These conservation equations are stated in the subsequent sections. 

\subsection{Momentum conservation for a hypersurface} \label{Sec:DiffHypSurf}
Now that the mass conservation for an EDH has been presented, similar steps can be used to obtain the momentum and energy conservation equations for an EDH.
%
The principal of momentum conservation states that the time rate of change of linear momenta of a material region is equal to the sum of forces acting on the region. This is mathematically presented as:
\begin{equation}
\frac{\text{d}}{\text{d}t} (\bf P) =  {\bf F}_{\text{surface}} + {\bf F}_{\text{body}} + {\bf \Pi}_{\text{mom}},
\end{equation}
where ${\bf P}$ is the total momentum in the material region, ${\bf F}_{\text{surface}}$ is the total surface force, ${\bf F}_{\text{body}}$ is the total body force, and ${\bf \Pi}_{\text{mom}}$ is the source of momentum. 

Further, writing in terms of field quantities and using the transport equation, we obtain the integral form of conservative momentum equation for the diffused region:
\begin{align}
\int_{m} \left\{\partial_t (\rho {\bf u})^m + \nabla^m \bm\cdot \left(\rho {\bf u} \otimes {\bf u}\right)^m = \nabla^m \bm \cdot {\bf T}^m + {\bf f}_{\text{body}}^m + \bm\pi_{\text{mom}}^m \right\} dx_{m}
\end{align}
Here, $ {\bf F}_{\text{surface}} = \int_m \left(\nabla^m \bm \cdot {\bf T}^m\right) dx_m$ and ${\bf T}^m$ is the stress tensor, ${\bf F}_{\text{body}}= \int_{m} \left( {\bf f}_{\text{body}}^m \right) dx_m$, and ${\bf \Pi}_{\text{mom}}= \int_{m}\left(\bm\pi_{\text{mom}}^m \right) dx_m$ . In order to ensure the generality of the momentum conservation equation, we do not substitute the constitutive relation for stress tensor. Which for a Newtonian fluid medium in $3$-D is, ${\bf T}^m =-p\bf I + \lambda(\nabla \bm\cdot \bf u)\bf I + \mu \left(\nabla \bf u + \nabla \bf u^T \right)$. Here, $p$ is the pressure, $\lambda$ is the bulk viscosity, and $\mu$ is the dynamic viscosity. 

Similar to the mass conservation, the gradient operator in the integral form of conservative momentum equation for diffuse hypersurface is decomposed into tangential and normal components. In addition,  the integral normal to the hypersurface is separated:
\begin{align}
\int_{m-1} \left\{\int_{n_1}^{n_2}  \left(\partial_t (\rho {\bf u})^m dn \right) + \int_{n_1}^{n_2} \nabla_{\hat{\bm s}}^{m-1} \bm\cdot \left(\rho {\bf u} \otimes {\bf u}\right)^m  dn + \int_{n_1}^{n_2} \nabla_{\hat{\bm n}}^{m} \bm\cdot \left(\rho {\bf u} \otimes {\bf u} \right)^m dn = \right. \notag \\ \notag
\left. \int_{n_1}^{n_2} \nabla_{\hat{\bm s}}^{m-1} \bm \cdot {\bf T}^m dn + \int_{n_1}^{n_2} \nabla_{\hat{\bm n}}^{m} \bm\cdot {\bf T}^m \cdot \hat{\bf n}dn + \int_{n_1}^{n_2} {\bf f}_{\text{body}}^m dn + \int_{n_1}^{n_2} \pi_{\text{mom}}^m dn \right\} dx_{m-1}
\end{align}
Using the Leibniz's rule and collapsing the diffused region by integrating in the direction normal to the EDH, the momentum conservation for an {\it actual} EDH is derived:
\begin{subequations}
\begin{align}
& \partial_t \left(\rho {\bf u}\right) ^{(m-1)'} +
\nabla_{\hat{\bm s}}^{m-1} \bm\cdot \left(\rho {\bf u} \otimes {\bf u}\right)^{(m-1)'} - \nabla_{\hat{\bm s}}^{m-1} \bm \cdot {\bf T}^{(m-1)'} - {\bf f}_{\text{body}}^{(m-1)'} -  \bm \pi_{\text{mom}}^{(m-1)'}   \\
& +\left[\!\left[ \left(\rho {\bf u} \otimes {\bf u} \right)^m \bm\cdot \hat{\bf n} \right]\!\right]  - \left[\!\left[{\bf T}^m \bm\cdot \hat{\bf n}\right]\!\right] \\
& + \left[\!\left[ \left(\rho {\bf u}\right)^{m}\frac{d (n)}{dt} \right]\!\right]_{n_1}^{n_2}  \\
& -  \left[\!\left[ \left(\rho {\bf u} \otimes {\bf u}\right)^{m} \bm \cdot \left( \nabla_{\hat{\bm s}}^{m-1} (n) \right) \right]\!\right]_{n_1}^{n_2} 
  -  \left[\!\left[ \left({\bf T}\right)^{m} \bm\cdot \left(\nabla_{\hat{\bm s}}^{m-1} (n) \right) \right]\!\right]_{n_1}^{n_2} = 0,
%
%
\end{align}
\end{subequations}

Finally, adjusting for the effects of the added homogeneous media, the momentum conservation for the {\it effective} hypersurface is presented as:

\begin{subequations}
\label{eq:EDH_Mom}
\begin{align}
& \overbrace{\partial_t \left(\rho {\bf u}\right) ^{(m-1)} }^{\substack{\text{rate of change of} \\ \text{Hypersurface} \\ \text{momentum}}}  
 ~+~\overbrace{\nabla_{\hat{\bm s}}^{m-1} \bm\cdot \left(\rho {\bf u} \otimes {\bf u}\right)^{(m-1)}}^{\substack{\text{Flux of} \\ \text{Hypersurface momentum}}}  
 ~ - ~\overbrace{\nabla_{\hat{\bm s}}^{m-1} \bm \cdot {\bf T}^{(m-1)}}^{\substack{\text{Force} \\ \text{at the boundary} \\ \text{of Hypersurface}}} \\
& - \overbrace{{\bf f}_{\text{body}}^{(m-1)}}^{\substack{ \text{hypersurface} \\ \text{ body force} }} -  \overbrace{\bm \pi_{\text{mom}}^{(m-1)}}^{\substack{\text{source of} \\ \text{hypersurface} \\ \text{momentum}}}  \\
& +   \overbrace{\left[\!\left[ \left(\rho {\bf u} \otimes ({\bf u} - {\bf v}) \right)^m \bm\cdot \hat{\bf n} \right]\!\right]_{n_0} }^{\substack{\text{jump in} \\ \text{bulk momentum}}} ~ - ~ \overbrace{\left[\!\left[{\bf T}^m \bm\cdot \hat{\bf n}\right]\!\right]_{n_0} }^{\substack{\text{jump in} \\ \text{bulk stress}}} 
  \\
&  ~-~ \overbrace{\left\{ \left[\!\left[ \left(\rho {\bf u} \otimes {\bf u}\right)^m \right]\!\right]_{n_0} ~ - ~ \left[\!\left[{\bf T}^m \right]\!\right]_{n_0} \right\} \bm\cdot \left( \nabla_{\hat{\bm s}}^{m-1} (n_0) \right)}^{\substack{\text{jump in bulk momentum and stress} \\ \text{due to variation in Divding Hypersurface location}}}  = 0
\end{align}
\end{subequations}

Here, there are two sets of additional terms to account for the flux due to spatial changes in the location of the EDH, equation \ref{eq:EDH_Mom}(d). One of those term accounts for the momentum flux $\left[\!\left[ \left(\rho {\bf u} \otimes {\bf u}\right)^m \bm\cdot \hat{\bf n} \right]\!\right]_{n_0} $ and the other for the jump in  bulk stress at these bounds $\left[\!\left[{\bf T}^m \bm\cdot \hat{\bf n}\right]\!\right]_{n_0}$. Finally, we derive the equations for the conservation of energy.

\subsection{Energy conservation}
The principal of conservation of total energy states that the time rate of change of total energy of a material region is equal to the net energy gained by the system from heat flux through the surface and work done on it. This is given as,
\begin{equation}
\frac{\text{d}}{\text{d}t} (E) = {Q} -  {W} + \Pi_{\text{energy}},
\end{equation}
where ${E}$ is the total energy in the material region, ${Q}$ is the total heat flux, ${W}$ is the total work done on the system, and ${\Pi}_{\text{energy}}$ is the source of energy. 

As previously done for mass and momentum equations, using the Reynold's transport theorem the equation for conservation of total energy is given as
\begin{align}
&\int_{m} \left\{
\partial_t \left(\rho \left(e + \frac{\bf u\cdot u}{2}\right) \right)^m + \nabla^m \bm\cdot \left(\rho {\bf u}\left(e + \frac{\bf u\cdot u}{2}\right) \right)^m = \right. \\ \notag &\left. \nabla^m \bm \cdot {\bf q}^m - \nabla^m \bm \cdot \left( {\bf T \bm \cdot  u} \right)^m + \pi_{\text{energy}}^m \right\} dx_m
\end{align}
Here, total energy $E= \rho e + ({\rho \bf u\cdot u})/{2}$,  $\rho e$ is internal energy,  $({\rho \bf u\cdot u})/{2}$ is kinetic energy, ${\bf q}$ is heat flux, $e$ is the specific internal energy, $Q=\int_m (\nabla^m \bm \cdot {\bf q}^m)dx_m$, $W=\int_m (\nabla^m \bm\cdot {\bf T \cdot u})^mdx_m$, and $\Pi_{\text{energy}}=\int_m (\pi^m_{\text{energy}})dx_m$. 

Since the steps are similar to that of mass and momentum conservation, we skip the intermediate steps and present the final relation for the conservation of energy for an effective EDH
\begin{subequations}
\begin{align}
&
\overbrace{\partial_t \left( \rho \left( e +  \frac{\bf u\cdot u}{2} \right) \right)^{(m-1)}}^{\substack{\text{rate of change of} \\ \text{EDH total energy}}} + 
\overbrace{\nabla_{\hat{\bm s}}^{(m-1)} \bm\cdot \left( \rho {\bf u} \left( e +  \frac{\bf u\cdot u}{2} \right) \right)^{(m-1)}}^{\substack{\text{convection of} \\ \text{EDH total energy}}}       \\ 
& - \overbrace{\nabla_{\hat{\bm s}}^{(m-1)} \bm \cdot {\bf q}^{(m-1)}}^{\substack{\text{gradient in} \\ \text{EDH heat}}}  +  \overbrace{\nabla_{\hat{\bm s}}^{(m-1)} \bm \cdot \left( {\bf T \bm \cdot  u} \right)^{(m-1)}}^{\substack{\text{gradient in work done} \\ \text{by EDH stress forces}}}   -  \overbrace{ \pi_{\text{energy}}^{(m-1)}}^{\substack{\text{source of} \\ \text{EDH} \\ \text{total energy}}}  \\
  & + \overbrace{\left[\!\left[  \left( \rho ({\bf u} - {\bf v})\left( e +  \frac{\bf u\cdot u}{2} \right) \right)^m \bm\cdot \hat{\bf n} \right]\!\right]}^{\substack{\text{jump in total energy} \\ \text{ of homogeneous media}}} - \overbrace{ \left[\!\left[ {\bf q}^m \bm\cdot \hat{\bf n} \right]\!\right]}^{\substack{\text{jump in} \\ \text{bulk heat}}}
 + \overbrace{\left[\!\left[ \left( {\bf T \bm \cdot  u} \right)^m \bm\cdot \hat{\bf n}\right]\!\right]}^{\substack{\text{jump in} \\ \text{bulk total work done}}} 
  \\ 
   & + \overbrace{\left\{\left[\!\left[  \left( \rho {\bf u} \left( e +  \frac{\bf u\cdot u}{2} \right) \right)^m \bm\cdot \hat{\bf n}  \right]\!\right] -  \left[\!\left[ {\bf q}^m \bm\cdot \hat{\bf n} \right]\!\right]
 + \left[\!\left[ \left( {\bf T \bm \cdot  u} \right)^m \bm\cdot \hat{\bf n}\right]\!\right] \right\}
  \bm\cdot \left( \nabla_{\hat{\bm s}}^{(m-1)} (n_0) \right)}^{\substack{\text{jump in bulk total energy, heat and work} \\ \text{due to variation in location of the EDH}}} = 0
\end{align}
\end{subequations}
Finally, with the conservation equations for an EDH presented, the equation of state for the EDH, followed by the boundary and initial conditions, are briefly discussed.

If in addition to velocity; density, temperature and pressure are also unknown, then an additional relation is required. This is given by the equation of state. In this paper it is assumed that the diffused region has an equation of state same as that of the homogeneous media and the EDH has an equation of state analogous to it. This might not necessarily always be the case. After deriving the above governing equations for an EDH it would not come as a surprise that the equation of state also needs to be re-derived by considering the collapse of the dimension.  This is beyond the scope of the current paper and is a topic that shall be addressed in future work.

Finally, in order to close the system, initial and boundary conditions are required.  Similar to homogeneous media, the field value of hypersurface quantities such as, velocity vector, pressure, density and temperature needs to be know at time $t=0$. In addition, the boundary conditions needs to be known for each of these hypersurface variables ($S_{\text{Hyp}}$, figure \ref{fig:Schematic-Arbitrary_Diff_Intf}). 

\subsection{Defining effective hypersurface quantities and identifying the location of the extended dividing hypersurface. \label{Sec:Def_HypQuant}} 

\subsubsection{Defining hypersurface quantities} 
\begin{figure}
\begin{minipage}{0.45\linewidth}
(a)
 \end{minipage}
  \begin{minipage}{0.45\linewidth}
(b)
 \end{minipage}\\
\begin{minipage}{0.45\linewidth}
\centerline{
 \includegraphics[width=1.0\textwidth]{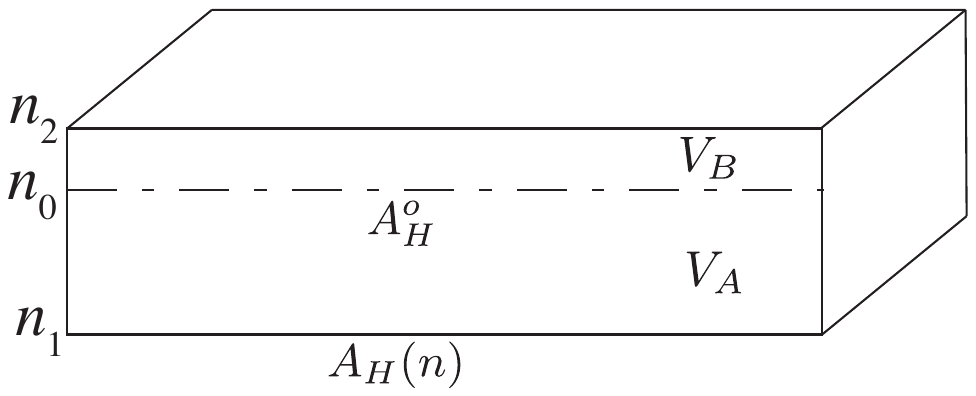}
 }
 \end{minipage}
 \begin{minipage}{0.45\linewidth}
\centerline{
 \includegraphics[width=1.0\textwidth]{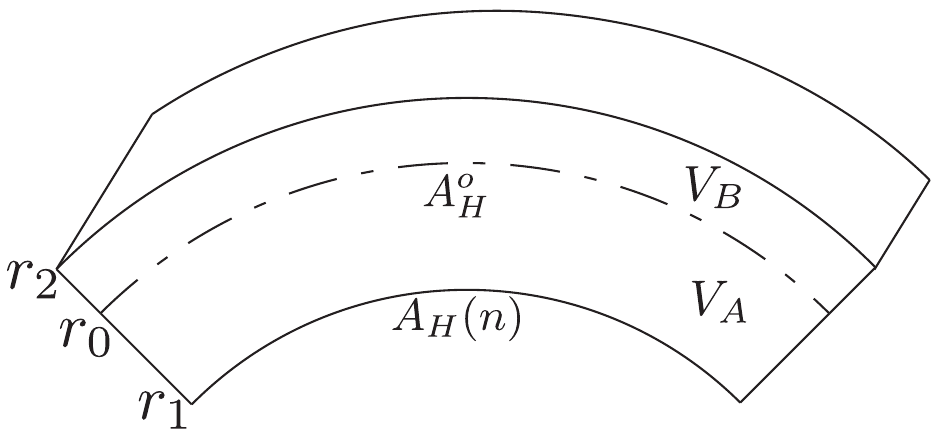}
 }
 \end{minipage}
   \caption{Schematic representation of a diffused region corresponding to a, a) planar and b) curved diffused region.}
\label{fig:Schematic-HyperArea}
\end{figure}

Following the derivation of the governing equations for the extended dividing hypersurface, the next step involves discussing the definitions of effective hypersurface quantities as well as the location of the EDH itself. When it comes to evaluating the hypersurface quantities, especially intensive quantities, it is done so by preserving the corresponding extensive quantities in a diffused region. That is, velocity is evaluated by preserving momentum, stress is computed by preserving the corresponding force, and temperature by preserving the heat energy. Therefore, for an arbitrary extensive or intensive parameter $\phi$, the hypersurface quantity is defined as
\begin{subequations}
\begin{equation}
\phi^{(m-1)}(s,b)_{\text{extensive}}  = \int_{n_1}^{n_2} \left(\phi ~A_{H}\right) dn - \underbrace{\left( \int_{n_1}^{n_0} \left(\phi_A~A_{H}\right) dn + \int_{n_0}^{n_2} \left( \phi_B~A_{H}\right) dn\right) }_{\text{Accounts for void}}.
\label{eq:SurfQty_Ext}
\end{equation}
\begin{equation}
\phi^{(m-1)}(s,b)_{\text{intensive}}  = \frac{\int_{n_1}^{n_2} \left(\phi ~A_{H}\right) dn - \left( \int_{n_1}^{n_0} \left(\phi_A~A_{H}\right) dn + \int_{n_0}^{n_2} \left( \phi_B~A_{H}\right) dn\right) }{A^o_{H}}.
\label{eq:SurfQty_Int}
\end{equation}
\end{subequations}
This definition is similar to that used to derive the governing equation for effective EDH. Here, $A_H(n)$ denotes the hyper-area bounding the diffused region, $\Sigma_A$ and $\Sigma_B$. It must be noted that $A_H(n)$ is not the cross-sectional area of the diffused region. $A_H^0$ corresponds to the hyper-area of the extended dividing hypersurface. Hence for example, referring to figure \ref{fig:Schematic-Arbitrary_Diff_Intf}, hyper-area for a $3$-D geometry is the area of the surface, for a $2$-D geometry is the perimeter of the line, and for $1$-D geometry is equal to 1, as tabulated in table \ref{tab:A_H}. In general, $A_H \neq A_H^0$, as seen in figure \ref{fig:Schematic-HyperArea}(b) but for geometries where the left and right boundaries of the diffused region, $S_{\text{Diff}}$, are parallel to each other, as in figure \ref{fig:Schematic-HyperArea}(a), $A_H = A_H^0 $.
\begin{table}
\begin{center}
\def~{\hphantom{0}}
\begin{tabular}{l*{2}{c}r} 
  & ~m~ & ~$A_{H}$~ \\[5pt]
& 1 &  1 \\
& 2 &  Perimeter \\
& 3 &  Area \\
& 4 &  Volume \\
& m &  cross section in $(m$-$1)$-D space \\
  \end{tabular}
  \caption{\label{tab:A_H} Hyper-area corresponding to $m$-dimensional space.}
\end{center}
\end{table}
%

In order to calrify the definition for a hypersurface quantity, the evaluation of hypersurface density will be explored. This evaluation considers a diffused region associated with both a curved and a planar hypersurface of unit depth, as depicted in figure \ref{fig:Schematic-HyperArea}. However, before delving into the specific evaluation, a conceptual overview will be presented on how the hypersurface quantity is computed.
\begin{align}
\rho^{(m-1)}(s,b)  &= \frac{\substack{\text{Mass of the}\\ \\ \text{diffused region}}  - \left( \substack{\text{Mass of media A} \\ \\ \text{extrapolated into the void}, V_A} + \substack{\text{Mass of media B} \\  \\\text{extrapolated into the void}, V_B}  \right) }{\text{Hyper-area of the hypersurface,}~\Sigma}
\end{align}
The definition proposed in this study differs from that used by \cite{SlatteryJC:64a} in terms of defining surface quantities based on extensive quantities. The underlying concept behind this approach is that the total mass remains constant between a diffused region and an extended dividing hypersurface (EDH). However, the magnitude of density, whether it is expressed as mass per unit volume, mass per unit area, or mass per unit length, may vary.
Similarly, the total force remains constant between a diffused region and a hypersurface, but the stress, whether it is measured as force per unit area or force per unit length, may differ. This is the rationale behind preserving the extensive quantities between the diffused region and the EDH, while the intensive quantities associated with the EDH are then derived from their corresponding extensive quantities.
%
%
Therefore, in the case of a planar hypersurface (figure \ref{fig:Schematic-HyperArea}(a)), $A_H(n) = A_H^0\equiv \text{constant}$, the hypersurface density is computed as:
\begin{align}
\rho^{(m-1)}(s,b)  &= \frac{\int_{n_1}^{n_2} \left(\rho(n) ~A_{H}(n)\right) dn - \left( \int_{n_1}^{n_0} \left(\rho_A~A_{H}(n)\right) dn + \int_{n_0}^{n_2} \left( \rho_B~A_{H}(n)\right) dn\right) }{A^o_{H}}, \notag \\
                     &= \int_{n_1}^{n_2} \rho(n) dn - \left[  \rho_A\left(n_0-n_1\right) +  \rho_B \left(n_2-n_0 \right) \right].
\end{align}
Here, it is assumed that the homogeneous media have a constant density. Hence, the extrapolated density adjacent to fluid A and B are $\rho_A$ and $\rho_B$, respectively. 
In the case of curved hypersurface (figure \ref{fig:Schematic-HyperArea}(b)), $A_H(n) =  R \Delta \theta$, where $R$ is the radius of curvature and $\Delta \theta$ is the angle across which the surface spans.
\begin{align}
\rho^{(m-1)}(s,b)  &= \frac{\int_{r_1}^{r_2} \left(\rho(n) ~A_{H}(n)\right) dn - \left( \int_{r_1}^{r_0} \left(\rho_A~A_{H}(n)\right) dn + \int_{r_0}^{r_2} \left( \rho_B~A_{H}(n)\right) dn\right) }{A^o_{H}},\\ \notag
			&= \frac{\int_{r_1}^{r_2} \left(\rho(n) ~R(n) \Delta \theta \right) dn - \left(  \rho_A(r_0^2 - r_1^2)\Delta\theta +  \rho_B \left( r_2^2 - r_0^2\right)\Delta\theta\right) }{r_0^2\Delta\theta},\\ \notag
                     &= \frac{\int_{r_1}^{r_2} \left(\rho(n) ~R(n) \right) dn - \left(  \rho_A(r_0^2 - r_1^2) +  \rho_B \left( r_2^2 - r_0^2\right)\right) }{r_0^2}
\end{align}
Hence, in the case of a {\it curved hypersurface, its density depends on the radius of curvature of the hypersurface}.
%
%
%
%


\subsubsection{Location of extended dividing hypersurface} 
From the definition of a hypersurface quantity it can be seen that its magnitude is directly dependent on the location of the EDH. The question that then remains is where in the diffused region is the EDH located? The answer is that the location of EDH is arbitrary. It can be any location as long as it is within the diffused region.
The EDH and its location is not unique to a given diffused region, rather it is dependent on (1) a pre-specified criteria and/or (2) the flow or fluid quantity the criteria is based on. The clarification of the aforementioned concepts is provided below.

Since the exact location of the EDH within the diffused region is arbitrary, additional information is required as an initial condition. This is provided in the form of criteria on which the location and consequently the EDH can be defined. The choice of criteria used to determine the location of EDH is based on convenience and does not effect the dynamics of the system as a whole, as long as the criteria are consistently applied throughout the set of equations describing the EDH. For example, Gibbs' proposed an EDH location based on surface tension and also one based on equimolar contributions of the two fluids \citep{GibbsJB:28a}. On the other hand Slattery defined the EDH to be located at a position where there is no effective mass of the EDH \citep{SlatteryJC:64a}. This dependence of the EDH location on the choice of criteria is no different than, for example trying to identify the center of a ship. The center of a ship, can be defined as the center-of-gravity, or the center of buoyancy, or the metacenter.

Secondly, it is a common assumption in the literature that the EDH location corresponding to its mass, momentum, and total energy are the same \citep{GibbsJB:28a,SlatteryJC:64a}. This is not always the case.  For example, if we choose the criteria that EDH is located where the effective mass, momentum, and total energy are all zero, then it will result in three different locations. Only for the case where the density distribution within the diffused region is a constant, will they all be at the same location. This dependence of the EDH location on a conserved flow quantity is analogous to the definition of the boundary layer thickness. For example, the thickness of the boundary layer can be based on the mass (displacement-thickness) or momentum (momentum-thickness), both of which are not always identical.

Considering that the EDH and its location are not unique to a diffused interface, the velocity of the EDH is not unique either. This is because the velocity of the EDH is computed as the time rate of change of its location, ${\bf v}=dn_0/dt$. It must be noted that ${\bf u}^{m-1}$ is not necessarily the same as ${\bf v} $.

\section{Test cases and numerical setup for validation}

In this paper, canonical example problems are used to validate the derived continuum model of an extended dividing hypersurface (EDH), which represents a fluid front. In order to validate it whenever possible, results are compared to the analytical solution of the Navier-Stokes equation. In instances where an analytical solution is unavailable, a more fundamental approach of molecular dynamics (MD) simulations is used to create a reference solution for comparison. MD simulations are used because of their ability to resolve the front.

In the case of examples using molecular dynamics simulations, bulk values from MD simulations at the boundaries of the diffused region are used as an input to the EDH equations. In the case of the example problem with an analytical solution, both the EDH equation and bulk equations are solved simultaneously. In this section, the problem geometry is described for each example, along with details of the simulations.

\subsection{Molecular dynamics simulations}

The LAMMPS package is used to perform molecular dynamics (MD) simulations \citep{PlimptonS:95a}. 

Here, the pairwise interaction of molecules, separated by a distance $r$, is modeled by the Lennard-Jones (LJ) potential
 \begin{equation}
 V^{LJ}=4\epsilon\left[\left(\frac{\sigma}{r}\right)^{12}-\left(\frac{\sigma}{r}\right)^{6}\right].
 \end{equation}
 Here, $\epsilon$ and $\sigma$  are the characteristic energy and length scales, respectively. The potential is set to zero for $r>r_c$, where $r_c$ is the cutoff radius. $r_c=2.5\sigma$, unless otherwise specified. 
 
  The temperature is maintained  using a Langevin thermostat with a damping coefficient of $\Gamma=0.1\tau^{-1}$, where $\tau=\sqrt{m\sigma^{2}/\epsilon}$ is the characteristic time and $m$ is the mass of the fluid molecule. As only 2D problems are simulated, the damping term is only applied to the $z$ direction to avoid biasing the flow.  The equation of motion of a fluid atom of mass $m$ along the $z$ component is therefore given as follows
 \begin{equation}
  m\ddot{z_i}=\sum_{j\neq i}\frac{\partial V{ij}}{\partial z{i}}-m\Gamma \dot{z_i} + \eta_i.
 \end{equation}
 Here, $\sum_{j\neq i}$ denotes the sum over all interactions and $\eta_i$ is a Gaussian distributed random force. The LJ coefficients and number density of the various cases simulated are listed in Table \ref{tab:test_case}. 
 The equations of motion were integrated using the Verlet algorithm \citep{VerletL:67a,TildesleyDJ:87a} with a time step $\Delta t=0.002\tau$. The molecular mass of individual atoms is 1. Hence mass density is equal to number density. Each specific problem simulated using MD is detailed next.
\subsubsection{Stationary fluids with varying miscibility}
\begin{figure}
\begin{minipage}{1.0\linewidth}
\centerline{
 \includegraphics[width=0.45\textwidth]{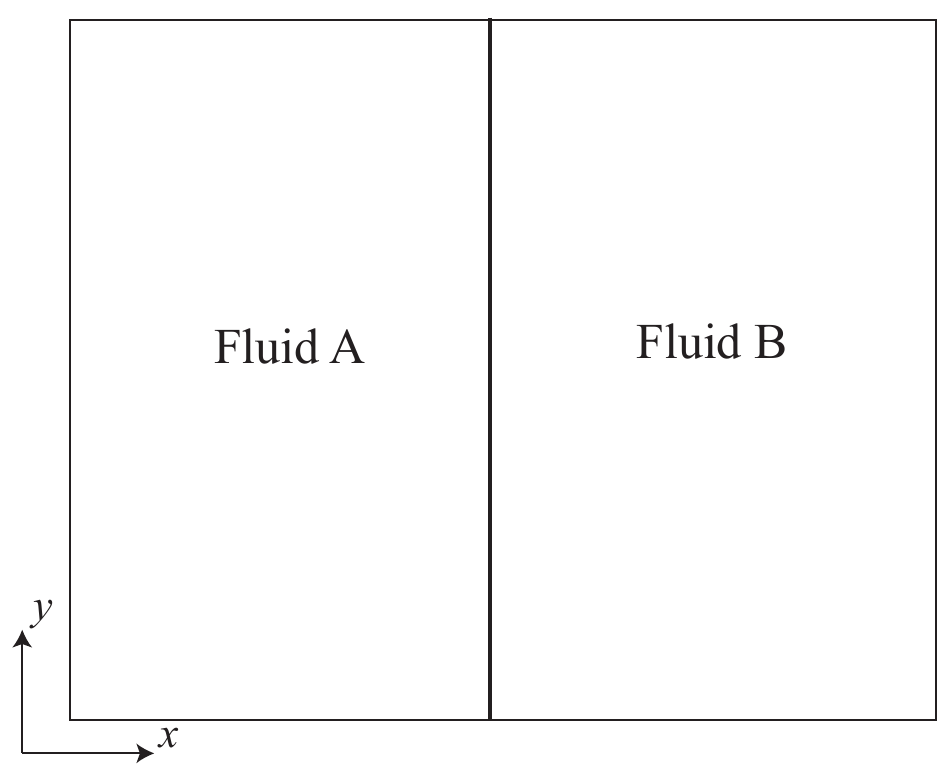}
 }
 \end{minipage}
  \caption{Schematic of two stationary fluids.}
\label{fig:Schematic-Stationary}
\end{figure}

Two stationary immiscible or partially miscible fluids at a constant temperature and pressure are simulated as shown in figure \ref{fig:Schematic-Stationary}. The domain is periodic in all directions. 

The domain size is $50 \sigma\times 30 \sigma\times 30 \sigma$. The Lennard Jones parameters for interatomic interactions and density of media used are listed in table \ref{tab:test_case}, Case 1-4.  It took the system $100~000$ steps to reach equilibrium. After which relevant data were extracted using spatial and temporal averaging. The data was averaged for $100~000$ steps and the bin size for spatial averaging was $0.5\sigma\times 30 \sigma\times 30 \sigma$. 

 \begin{table}
   \begin{center}
 \def~{\hphantom{0}}

   \begin{tabular}{cccccccccc}
        & \multicolumn{3}{c}{\underline{fluidA-fluid A}} & \multicolumn{3}{c}{\underline{fluidB-fluid B}} & \multicolumn{3}{c}{\underline{fluid A-fluid B}}\\[3pt]
     Case & ~$\epsilon^{AA}/\epsilon$~ & ~$\sigma^{AA}/\sigma$~ & ~$\rho^*$~ & ~$\epsilon^{BB}/\epsilon$~ & ~$\sigma^{BB}/\sigma$~ & ~$\rho^*$~ & ~$\epsilon^{AB}/\epsilon$~ & ~$\sigma^{AB}/\sigma$~ & ~$r_c/\sigma$~\\
       1~ &~$1.00$~ & ~$1.00$~ & ~$0.78$~  &~$1.00$~ & ~$1.00$~ & ~$0.78$~  &~$0.01$~ & ~$4.00$~ & ~$7.00$~\\ 
       2~ &~$1.00$~ & ~$1.00$~ & ~$0.82$~  &~$1.00$~ & ~$1.00$~ & ~$0.82$~   &~$0.01$~ & ~$1.00$~ & ~$2.50$~\\ 
       3~ &~$1.00$~ & ~$1.00$~ & ~$0.78$~  &~$1.00$~ & ~$1.00$~ & ~$0.78$~    &~$0.01$~ & ~$1.00$~ & ~$2.50$~\\
       4~ &~$0.75$~ & ~$1.25$~ & ~$0.18$~  &~$1.50$~ & ~$1.00$~ & ~$0.78$~    &~$0.75$~ & ~$1.25$~ & ~$2.50$~\\
       5~ &~$0.75$~ & ~$1.25$~ & ~$0.18$~  &~$1.50$~ & ~$1.00$~ & ~$0.78$~    &~$0.50$~ & ~$1.50$~ & ~$4.00$~\\
       6~ &~$1.00$~ & ~$1.00$~ & ~$0.81$~  &~$1.00$~ & ~$1.00$~ & ~$0.81$~  &~$0.20$~ & ~$3.00$~ & ~$5.00$~\\ 
   \end{tabular}
       \caption{List of different test cases. Here, $\epsilon$ and $\sigma$  are the characteristic energy and length scales, respectively. $\rho^*$ is the number density.}
   \label{tab:test_case}
   \end{center}
 \end{table}

\subsubsection{Stratified flow through a converging diverging section}
\begin{figure}
\begin{minipage}{1.0\linewidth}
\centerline{
 \includegraphics[width=0.55\textwidth]{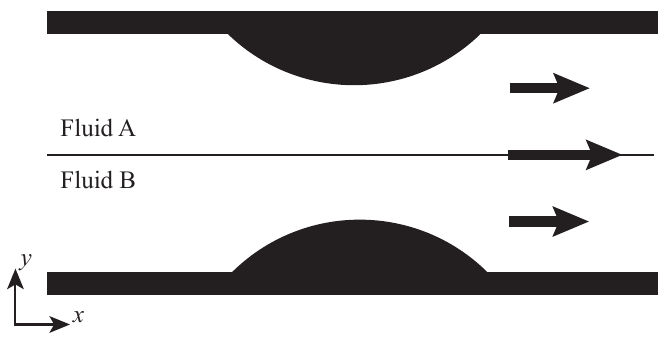}
 }
 \end{minipage}
 \\
 \caption{Schematic of a stratified flow through a converging-diverging channel.}
 \label{fig:Schem_ConvgDivg}
 \end{figure}

In this test case, a varying cross sectional area is modeled by a converging-diverging section in the channel, with a periodic domain in the $x$ and $z$ directions. Two immmiscible fluids are subjected to a constant body force of $0.01 \epsilon\sigma^{-1}$in the $x$ direction. The schematic of the problem is presented in figure \ref{fig:Schem_ConvgDivg}.

The domain size is $127.43 \sigma\times 55.90 \sigma\times 27.58 \sigma$. 
Each wall is comprised of at least two layers of molecules oriented along the (111) plane of a face centered cubic (fcc) lattice, with the molecules fixed to their respective lattice sites. The wall number density is $3.24\sigma^{-3}$. The LJ parameters for wall-fluid interactions for both the fluids are $\epsilon^{wf}=0.2\epsilon$ and $\sigma^{wf}=2.0\sigma$, with a cut-off radius of $r_c=5\sigma$. The LJ parameter for fluid-fluid interactions are given by Case 6 in table \ref{tab:test_case}.
It took the system $50~000$ steps to reach equilibrium. The extracted data was averaged  for $10~000~000$ steps and the bin size for spatial averaging was $\approx 1.0\sigma\times 1.0 \sigma\times 27.58 \sigma$. 
 

\subsubsection{Shock tube problem}
The canonical shock tube problem considered here is a long tube which is closed at both ends. A diaphragm separates the region of high-pressure fluid on the right from the region of low-pressure fluid on the left. When the diaphragm is broken at $t=0$ a shock wave propagates into the low pressure region, towards the left, and an expansion wave propagates towards the right, as illustrated in the schematic in figure \ref{fig:Schematic-ShockTube}. The diaphragm is simulated in MD by two layers of wall atoms which are removed at time $t=0$ to initiate the propagation of shock and expansion waves.
\begin{figure}
\begin{minipage}{1.0\linewidth}
\centerline{
 \includegraphics[width=0.55\textwidth]{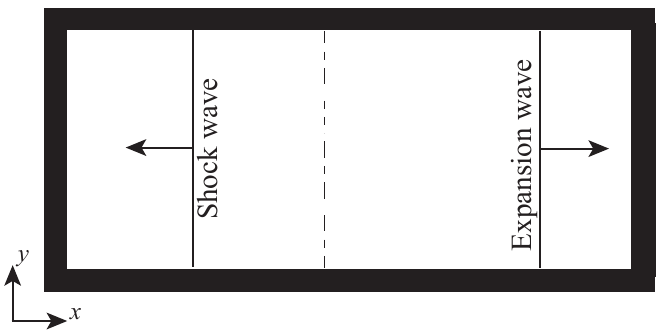}
 }
 \end{minipage}
 \\
 \caption{Schematic of a shock tube problem.}
 \label{fig:Schematic-ShockTube}
 \end{figure}

The domain size is $298.11 \sigma\times 231.92 \sigma\times 210.09 \sigma$. 
Each wall is comprised of at least two layers of molecules oriented along the (111) plane of a face centered cubic (fcc) lattice, with the molecules fixed to their respective lattice sites. The wall number density is $3.24\sigma^{-3}$. The LJ parameters for wall-fluid interactions for both the fluids are $\epsilon^{wf}=0.1\epsilon$ and $\sigma^{wf}=1.0\sigma$. 
It took the system $100~000$ steps to reach equilibrium. After which the membrane dividing the fluid in two different state is removed by deletion and the spatial and temporal average was performed. The data was averaged  for $500$ steps and the bin size for spatial averaging was $\approx 1.0\sigma\times 231.92 \sigma\times 210.09 \sigma$. 

\subsection{Example with an analytical solution}

\subsubsection{Bubble dynamics}
The last example problem looks at the evolution of a bubble when subjected to a non-equilibrium initial condition. The non-equilirbium condition is initiated by a high pressure at infinity. 
The radial evolution of an unstable bubble in an unbounded liquid is given by the Rayliegh-Plesset equation \citep{Rayleigh:1917a,PlessetMS:49a,ProsperettiA:04a}. In this paper, the Rayleigh-Plesset equation is modified by considering the physical interface having a finite mass and thickness. 

\begin{figure}
\center
\begin{minipage}{0.85\linewidth}
\centerline{
 \includegraphics[width=0.55\textwidth]{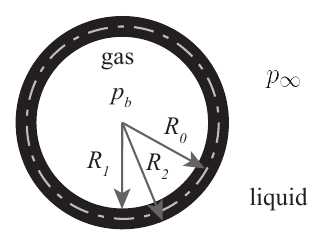}
 }
 \end{minipage}
 \\
 \caption{Schematic of the unsteady bubble dynamics problem.}
 \label{fig:Schematic-ShockTube}
 \end{figure}

In this example problem, the {\it actual} hypersurface description of the EDH is used.
The hypersurface is considered to be located ($R$) at the geometric mean (center) of the bounds of the diffused interface. The locations of the inner and outer bounds are given as $R_{1}$ and $R_{2}$. Hence, $R = (R_{1}+R_{2})/2$. Since the physical thickness of the interface needs to be accounted for, the Rayleigh-Plesset equation is modified as,
\begin{align}
p_{b,(r=R_1)} - p_{\infty} = \frac{2\gamma}{R_0} + 4\mu\frac{u_{(r=R_{2})}}{R_{2}^3} +  R_{out}\frac{\partial u_{(r=R_{2})}}{\partial t} - \frac{\rho}{2} u_{(r=R_{2})}^2 
\end{align}
Here, using ideal gas law, the pressure within the bubble, $p_b$, is computed for a radius of $R_1$. In addition, it is assumed that hypersurface (interface) momentum, $ (\rho u)^s$ does not change with time.


In order to compute hypersurface quantities, such as hypersurface mass, it is considered that the density sequence (function), describing the density within the interface, is a 4th-order polynomial. In addition, it is assumed that the magnitude and the gradient of density in the radial direction are continuous at the bounds of the interfacial region. This is used to evaluate interfacial quantities analytically. 

The evolution of bubble radius is found by simultaneously solving the mass conservation of the hypersurface and the Rayleigh-Plesset equation. This modified system of equations is solved numerically. It must be noted as a means of validating and demonstrating the implementation of the EDH equations, only the hypersurface mass conservation is used for this example. The momentum conservation will be incorporated in future work.


The non-dimensional parameters and initial conditions chosen for this parametric numerical study are $\gamma=1.4$ coefficient of polytropic expansion, the viscosity of the liquid $\mu_l=4000$, the density of the liquid $\rho_l=10$, the density of gas within the bubble $\rho_g=0.01\rho_l$, surface tension $\sigma = 3$, pressure at infinity is $100\times P_{in}$, the initial radius of the bubble $R_0=1$ initial thickness of interface $1/20R_0$, and initial pressure inside the bubble $10^6$. 




\section{Results and discussion}

The primary objective of this paper is to expand upon the concept of a dividing surface initially introduced by Gibbs. The aim is to generalize this concept to encompass any fluid and flow front, going beyond just the phase interface. In order to achieve this goal, the authors examine a series of canonical problems involving fluid and/or flow fronts.

These problems serve four main purposes:
\begin{enumerate}
\item They illustrate that the extended dividing hypersurface (EDH) has the capability to accurately capture the dynamics of not only phase or material interfaces but also other types of fluid and flow fronts, specifically shock front (physical front) and vortex sheet (apparent front).
\item They emphasize that the distribution of monotonicity within a front, as commonly described in literature, is just one of several possible functionalities that the EDH can capture.
\item They demonstrate the relationship between the flux of $m$-dimensional quantities and the $m$-1 dimensional quantities (hypersurface quantities), highlighting how this coupling can lead to hypersurface dilatation even in incompressible hypersurface flows. This finding contradicts deductions made from continuity equations for a bulk fluid.
\item They underscore the importance of acknowledging the mass of the front and consequently demonstrating its impact on the dynamics of the front. 
\end{enumerate}

\subsection{Stationary fluids with varying miscibility}\label{subsec_1}

We start with the simplest example of two stationary fluids adjacent to each other, see figure \ref{fig:Schematic-Stationary}. As a result of the fluids being stationary the EDH has no dynamics. Hence, the only non-zero quantity associated with the EDH are the thermodynamic quantities. Furthermore, since we consider a system with a planar EDH and a constant temperature, the only thermodynamic quantity that is discontinuous across the EDH is density. As we have done through the course of this paper, looking at density or mass of the EDH serves as the best starting example to understand hypersurface quantities.

\begin{figure}
\begin{minipage}{0.5\linewidth}
a) negative effective surface mass
\end{minipage}
\begin{minipage}{0.5\linewidth}
b) positive effective surface mass
\end{minipage}\\
\begin{minipage}{0.5\linewidth}
\centerline{
 \includegraphics[width=1.1\textwidth]{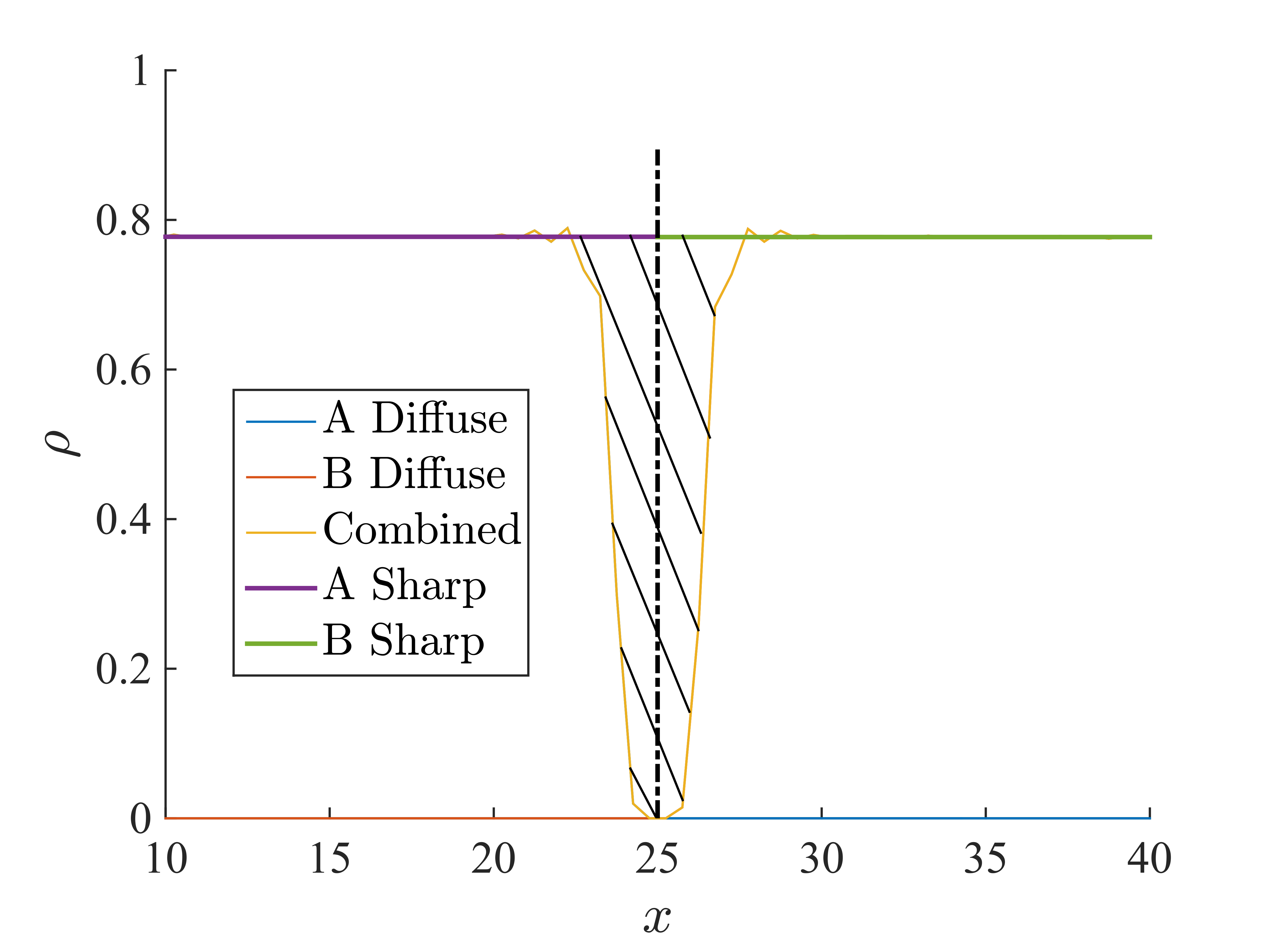}
 }
 \end{minipage} 
 \begin{minipage}{0.5\linewidth}
\centerline{
 \includegraphics[width=1.1\textwidth]{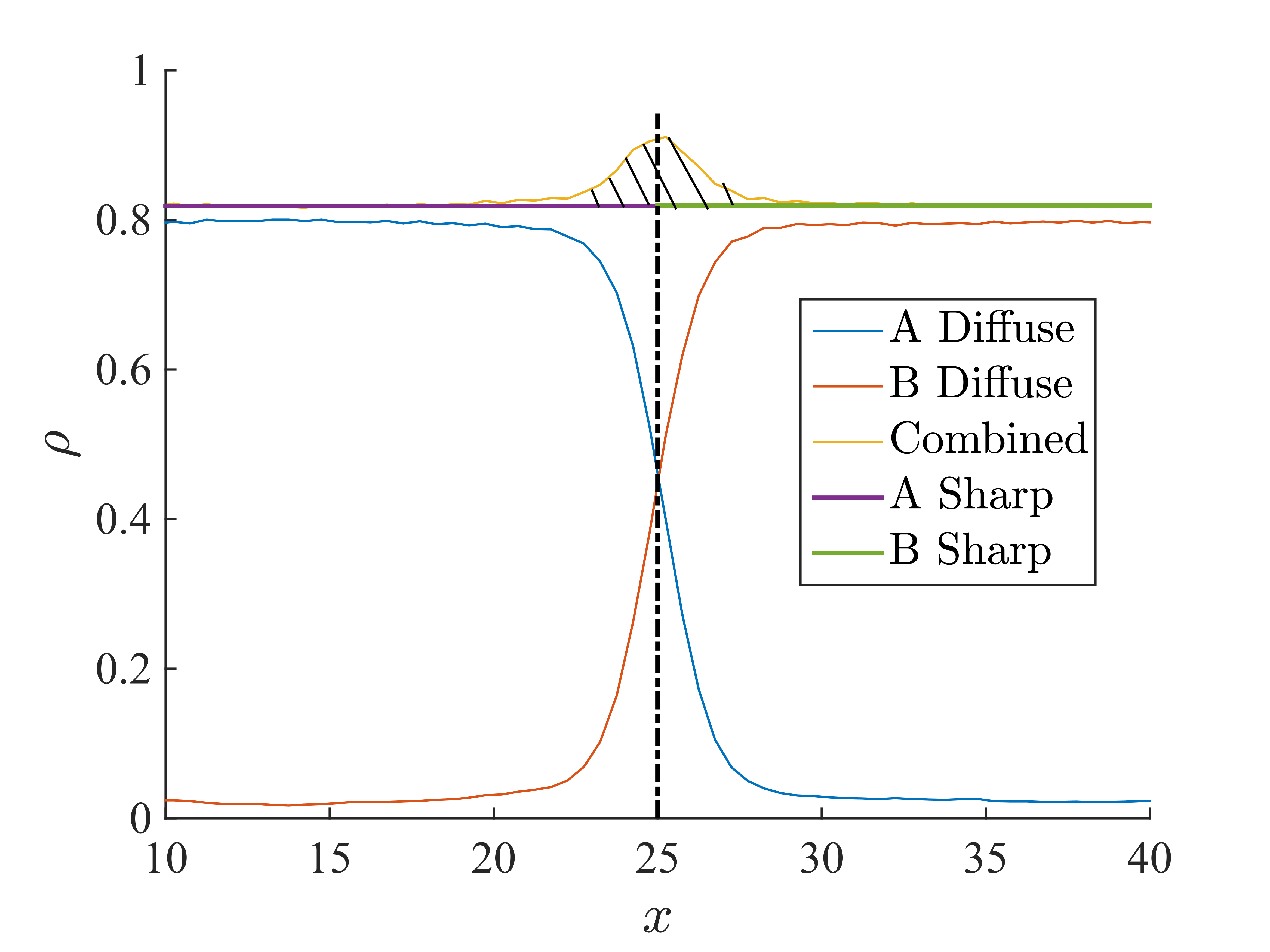}
 }
 \end{minipage} \\
 \begin{minipage}{0.5\linewidth}
c) zero effective surface mass
\end{minipage}
 \begin{minipage}{0.5\linewidth}
d) zero effective surface mass
\end{minipage}\\
 \begin{minipage}{0.5\linewidth}
\centerline{
 \includegraphics[width=1.1\textwidth]{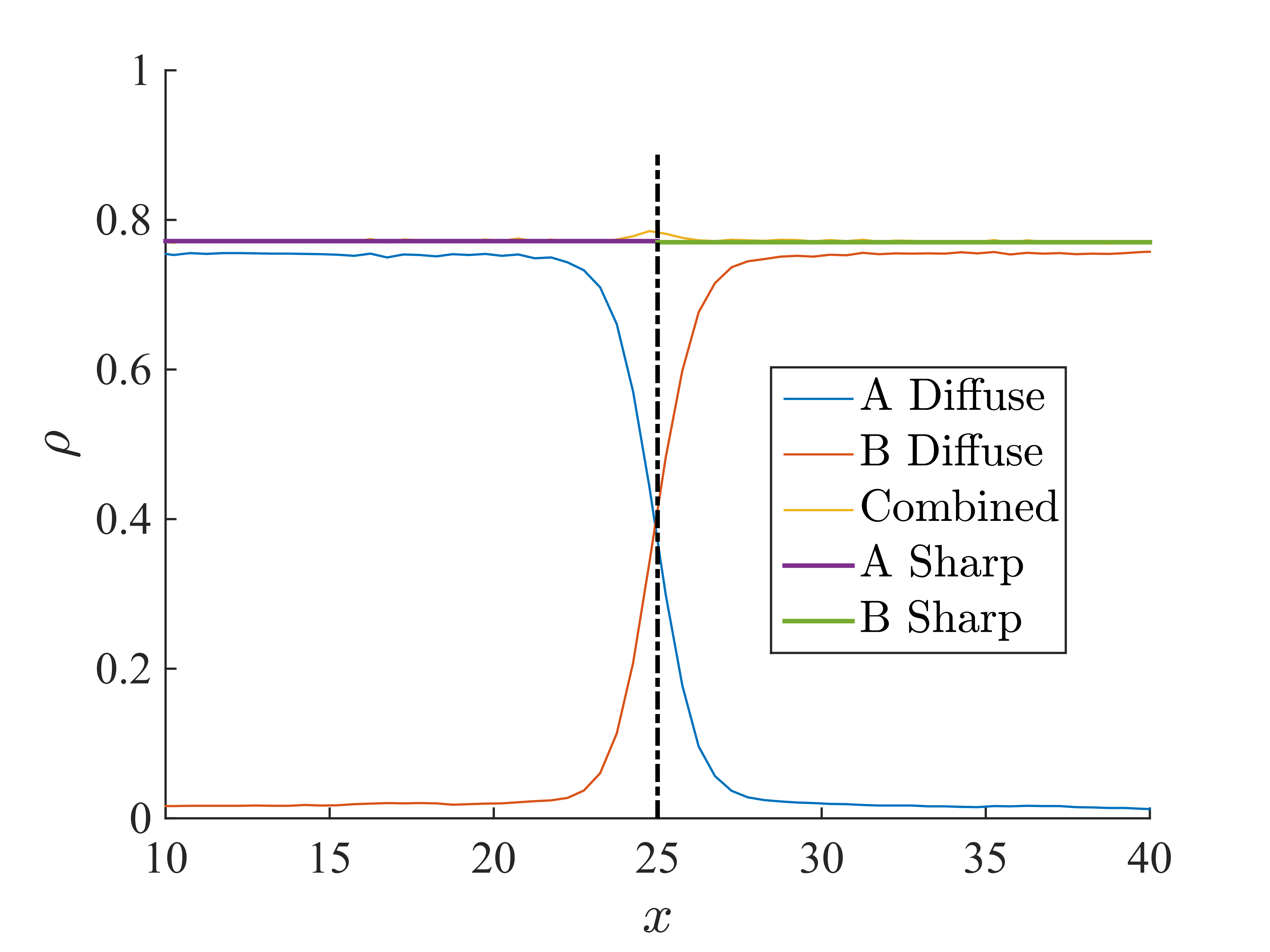}
 }
 \end{minipage} 
  \begin{minipage}{0.5\linewidth}
\centerline{
 \includegraphics[width=1.1\textwidth]{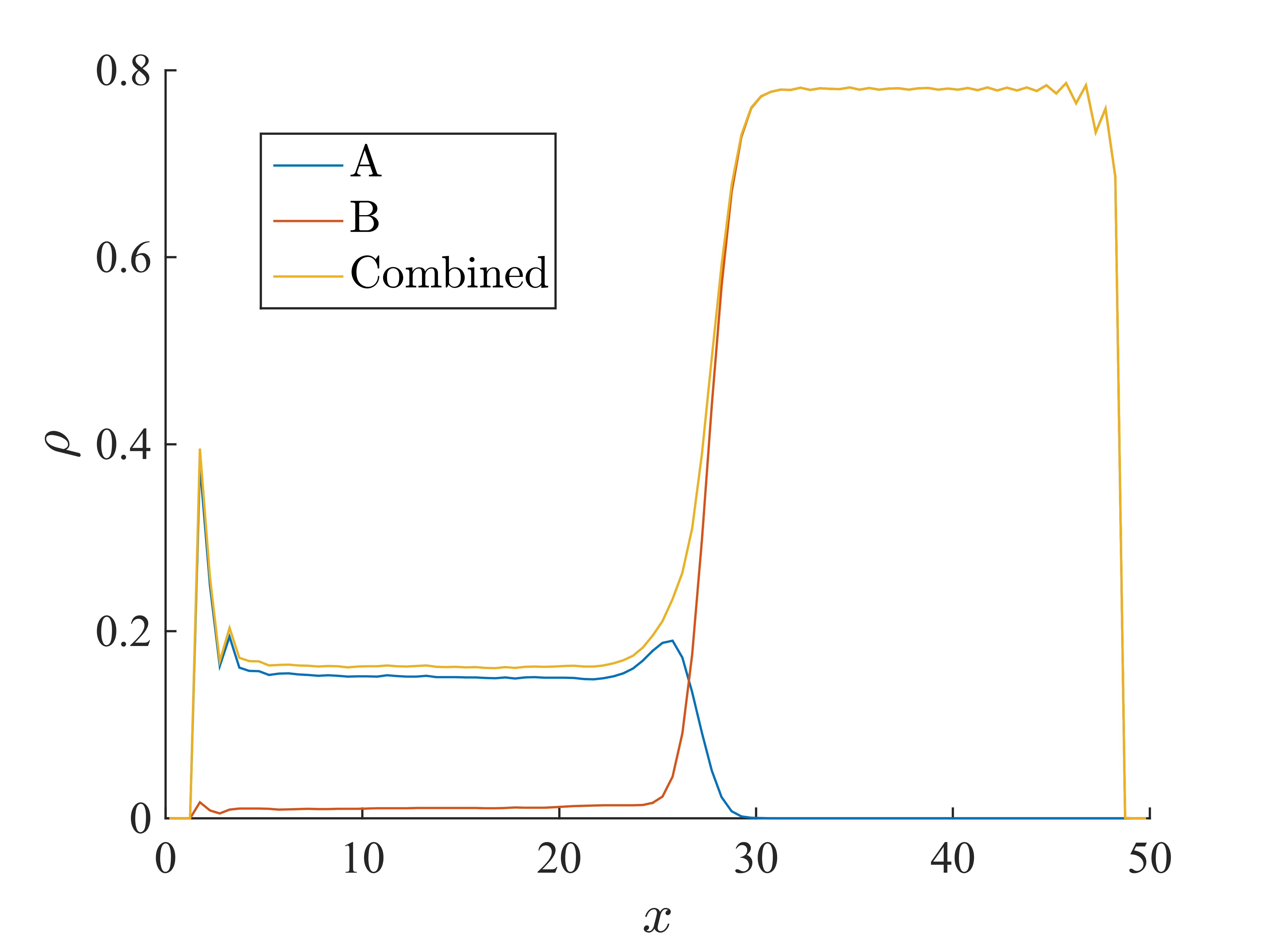}
 }
 \end{minipage} 
  \caption{Compares the density distribution of individual fluids along with the total density, across the diffused region. Figures a)-c) consider the two fluid to have identical density and individual density profiles showing a monotonic behavior.  Figure d) considers a case with the two fluids having two different densities, with one of the fluid showing a non-monotonic behavior.}
\label{fig:Monotonic}
\end{figure}

Different test cases with varying density ratios and miscibilities are presented. This helps us demonstrate various nuances of hypersurface quantities and the common assumptions (explicit or implicit) made about them in the literature. We first put to test the assumption that the density profile is monotonic across the diffused region and the local value being  never greater or less than the value of bulk densities as was shown in figure \ref{fig:SchemCollapse}. This is especially true, when numerically modeling a material or a phase interface (front-tracking, level-set, phase field methods \citep{ProsperettiA:07a}). Referring to figure \ref{fig:Monotonic}, it is seen that in the first two cases the individual density distribution is monotonic, but the combined densities are not. The behavior of both individual and combined density profiles in case (c), is closest to the monotonic assumption made in literature, but it also happens to be a trivial solution. As for the last example, of a stationary gas next to a liquid, case (d), the local density distribution of fluid A, is also not monotonic. 

Next, we look at the actual and effective hypersurface mass, which is evaluated as per the definition presented in equation \eqref{eq:SurfQty_Ext}. 
Here, we consider the EDH to be located where the density profiles of the two fluids intersect, as seen in figure \ref{fig:Monotonic} and denoted by a dashed line. Bold line shows the extrapolated bulk density of individual fluids. Hence, the effective hypersurface mass, or as Gibbs calls it, the excess mass, is given by the hatched region in the figure.  It can be seen that figures (a), (b) and (c) give a negative, positive and zero effective mass, respectively. This is contrary to what is commonly assumed in literature. It is commonly assumed that the interface has no mass, when considering perfectly immiscible fluids in continuum simulations \citep{ProsperettiA:07a}, whereas results from MD, presented here, show that it can in fact have both a positive and negative effective mass. 
The hypersurface density is not separately discussed, because as mentioned in section \ref{Sec:Def_HypQuant}, it is derived from hypersurface mass.

 In the static case, with no curvature, an argument can be made that because the relative thickness of the diffused region is negligible compared to the length scale of the homogeneous media, the zero hypersurface mass is  a good assumption. While this is true for effective mass, the same cannot be always said for effective momentum, energy or stress. For instance, effective hypersurface linear stress gives the hypersurface pressure, which is nothing but the mechanical surface tension, which we know not to be negligible. Hence, from MD simulation results and definition of hypersurface quantities presented in section sec. \ref{Sec:Def_HypQuant} a case is made that the assumptions:
 \begin{enumerate}
 \item that distribution across a diffused region is monotonic,
 \item the local value within this region lies always within the range of the corresponding bulk values, and
 \item the effective hypersurface mass is zero,
 \end{enumerate}
 are not always true.

\subsection{Stratified flow through a converging diverging section}
In the previous example since there was no gradient along the EDH, there were no internal dynamics or viscous stresses in the EDH. In this section a 2D stratified fluid flow through a section with varying cross sectional area is considered. The varying cross-section of the channel results in accelerating the flow. This example is used to demonstrate the relationship between the mass flux of the bulk fluid into the EDH and the hypersurface dilatation. This relation is  directly obtained from the mass conservation equation for the EDH, and as such also acts as its validation. In addition, we discuss the recent finding by \cite{Mohseni:20a}, suggesting the deviation of mechanical and thermodynamic surface tension in the presence of hypersurface dilatation. 
%
%

In the case of a stratified Couette flow the list of relevant assumptions made are as follows:
\begin{enumerate}
\item Steady state, no time rate of change of hypersurface quantities.
\item No source of hypersurface mass or momentum.
\item No hypersurface body force.
\item Location of the bounds of the boundary layer (or width) does not change with time and space. 
\item Since we are considering a periodic or 2D problem, there is no surface shear $T_{sn}^{(m-1)'}=0$.
\item We assume that there is no surface dilatation, $\nabla^{m-1} \bm\cdot u^{(m-1)'}=0$.
 \item There is no surface gradient in surface pressure.
\end{enumerate}
Applying these assumptions, the governing equations for the {\bf actual extended dividing hypersurface} become:\\
  
  {\it Mass conservation}, 
\begin{align}
\nabla_{\hat{\bm s}}^{m-1} \bm \cdot (\rho{\bf u})^{(m-1)'}  ~ +  ~  \left[\!\left[(\rho{\bf u})^m \bm\cdot \hat{\bf n} \right]\!\right]_{n_1}^{n_2} =  0 
\label{eq:Mass_ConvgDivg_act}
\end{align}
 
 {\it Momentum conservation},
 \begin{align}
\nabla_{\hat{\bm s}}^{m-1} \bm\cdot \left(\rho {\bf u} \otimes {\bf u}\right)^{(m-1)'} - \nabla_{\hat{\bm s}}^{m-1} \bm \cdot {\bf T}^{(m-1)'}
 + \left[\!\left[ \left(\rho {\bf u} \otimes {\bf u} \right)^m \bm\cdot \hat{\bf n} \right]\!\right]_{n_1}^{n_2}        - \left[\!\left[{\bf T}^m \bm\cdot \hat{\bf n}\right]\!\right]_{n_1}^{n_2}   = 0.
\end{align}

 Similarly, the conservation equations for the {\bf effective extended dividing hypersurface} reduce to:\\
 
 {\it Mass conservation}, 
\begin{align}
\nabla_{\hat{\bm s}}^{m-1} \bm \cdot (\rho{\bf u})^{(m-1)}  ~ +  ~  \left[\!\left[(\rho{\bf u})^m \bm\cdot \hat{\bf n} \right]\!\right]_{n_0} =  0 
\label{eq:Mass_ConvgDivg_eff}
\end{align}
 
 {\it Momentum conservation},
 \begin{align}
\nabla_{\hat{\bm s}}^{m-1} \bm\cdot \left(\rho {\bf u} \otimes {\bf u}\right)^{(m-1)} - \nabla_{\hat{\bm s}}^{m-1} \bm \cdot {\bf T}^{(m-1)}
 + \left[\!\left[ \left(\rho {\bf u} \otimes {\bf u} \right)^m \bm\cdot \hat{\bf n} \right]\!\right]_{n_0}       - \left[\!\left[{\bf T}^m \bm\cdot \hat{\bf n}\right]\!\right]_{n_0}  = 0.
\end{align}

\begin{figure}
\begin{minipage}{1.0\linewidth}
\centerline{
 \includegraphics[width=0.65\textwidth]{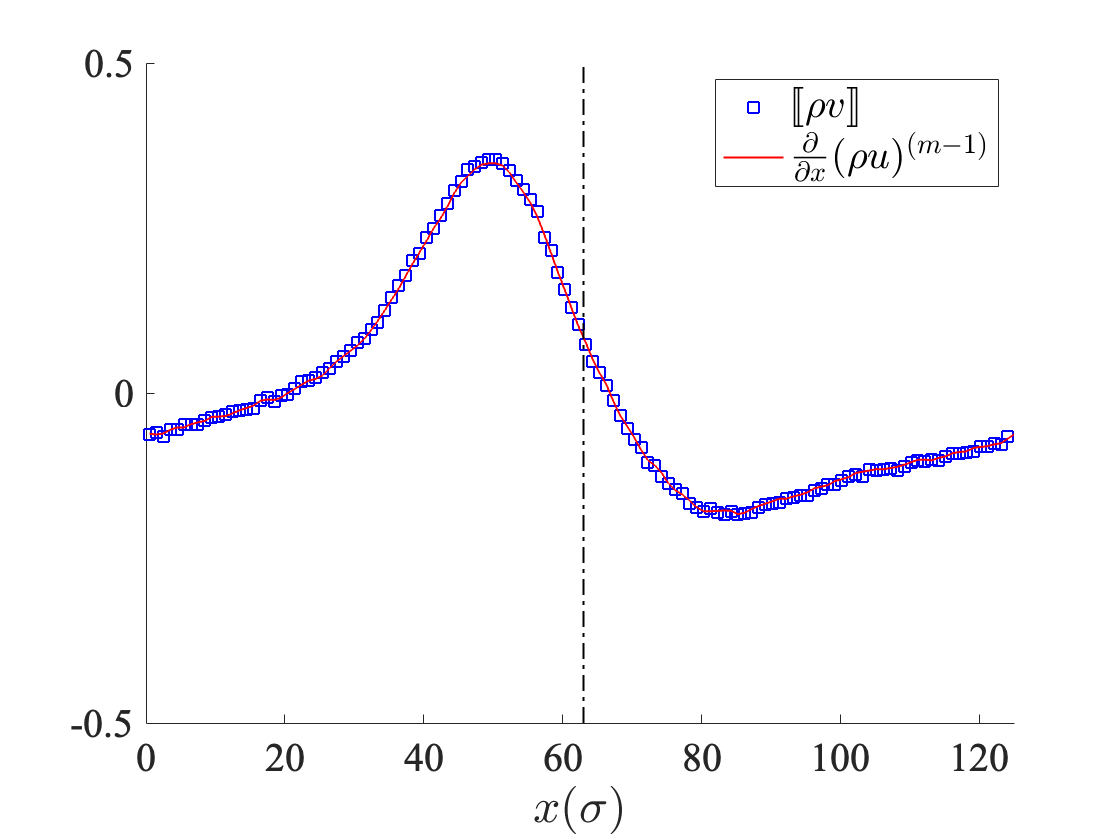}
 }
 \end{minipage}
 \\
  \caption{Comparing of hypersuface mass flux within the EDH (or surface dilatation) with the jump in mass flux of the homogenous bulk media across the EDH. These results from MD simulation help confirm the EDH mass conservation equation, eq. \ref{eq:Mass_ConvgDivg_eff}. The black dash-dot line represent the location of the throat of the channel.}.
\label{fig:Entrainment_SurfaceFlux}
\end{figure}

 
Recalling that the conservation of mass for an EDH is obtained by collapsing the dimension and integrating the conservation equation of the homogeneous media in the normal direction. When the dimension is collapsed, the term $\bm \nabla \bm \cdot {\bf u}$, that represents the flux of mass from one region to another, transforms into a relation that represents the flux of mass from a higher dimension to a lower dimension, given by the jump, $\left[\!\left[(\rho{\bf u})^m \bm\cdot \hat{\bf n} \right]\!\right]$. While for a bulk fluid, in 3D space, the incompressible condition ($ D  \rho / Dt = 0$) results in the flow having no dilatation ($\bm \nabla \bm \cdot {\bf u} = 0$), for an EDH, in 2D space, there can be surface dilatation even though the surface fluid is incompressible. This is because the surface dilatation can still be caused by the bulk mass flux into or out of the EDH. This fact which is represented in equations  \eqref{eq:Mass_ConvgDivg_act} and \eqref{eq:Mass_ConvgDivg_eff}, is further validated by looking at MD results depicting the net mass flux from the homogeneous media in 3D into or out of the EDH (represented by $\left[\!\left[(\rho{\bf u})^m \bm\cdot \hat{\bf n} \right]\!\right]$ ) and the hypersurface mass flux within the EDH (given by $\nabla_{\hat{\bm s}}^{m-1} \bm \cdot (\rho{\bf u})^{(m-1)}$), as shown in figure \ref{fig:Entrainment_SurfaceFlux}.
 %
%

\begin{figure}
\begin{minipage}{1.0\linewidth}
\centerline{
 \includegraphics[width=0.65\textwidth]{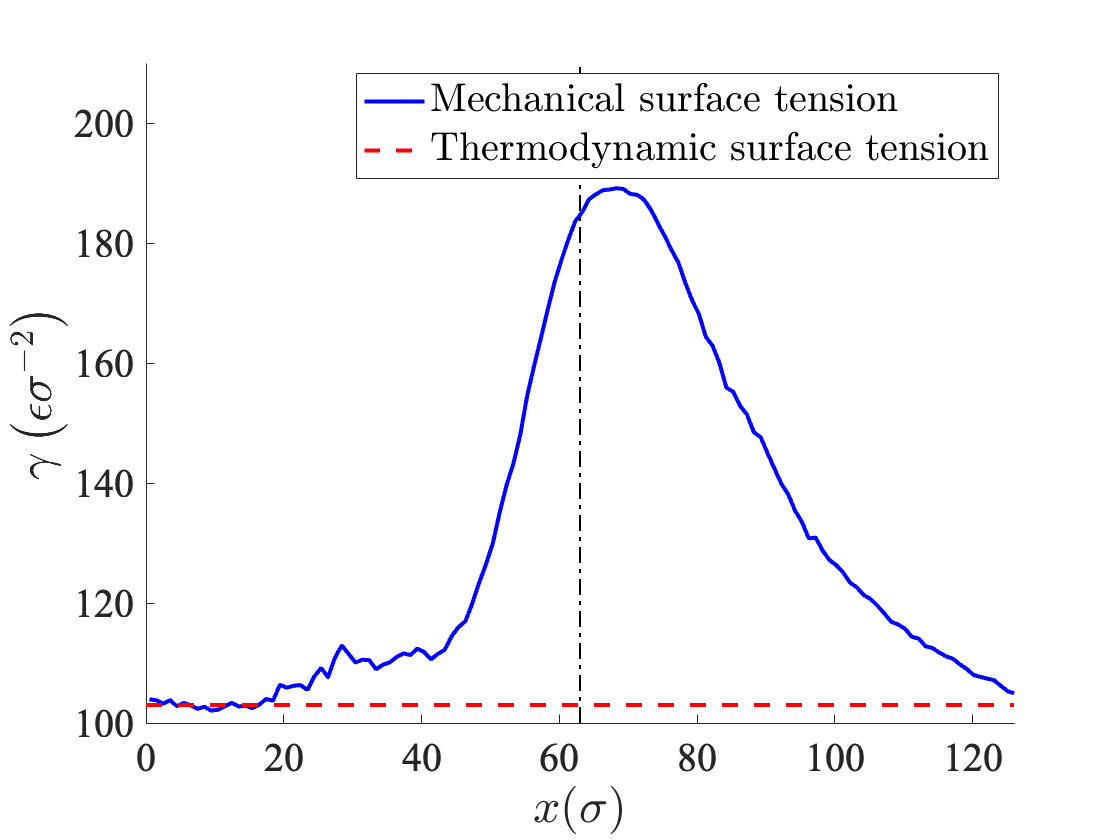}
 }
 \end{minipage}
 \\
  \caption{Deviation of local mechanical surface tension value from the thermodynamic surface tension. The black dash-dot line represent the location of the throat of the channel.} 
\label{fig:Mech_Thermo_SurfTens}
\end{figure}

Another important consequence of the surface dilatation is that the mechanical surface tension is no longer equal to the thermodynamics surface tension. Recently, \citep{Mohseni:20a} presented that analogous to the relation between mechanical and thermodynamic pressure, mechanical and thermodynamic surface tension are related as $\gamma_m = \gamma_t + (\lambda_s + \mu_s)\nabla_{(m-1)} \cdot u^{(m-1)}$, where $\gamma_m$ is the mechanical surface tension, $\gamma_t$ is the thermodynamic surface tension, $\lambda_s$ and $\mu_s$ is the first and second surface viscosities. The major difference, is the fact that in the EDH (2D), hypersurface dilatation can also be caused by mass flux of bulk homogeneous fluid into or out the EDH. In figure \ref{fig:Mech_Thermo_SurfTens} we plot the distribution of local mechanical surface tension along the length of the EDH, for the case where the force acting on the fluid is $f=0.02$. The deviation of the mechanical surface tension from the constant value of a thermodynamic surface tension is clearly visible.


\subsection{Blasius boundary layer}
So far all test cases have been where material/phase interfaces have been represented as an EDH. Here, we demonstrate that the EDH can be used to model a viscous boundary or shear layer. This is done by showing that the vortex-entrainment sheet, recently introduced by \cite{Mohseni:19a, MohseniK:17a} is just a limiting case of the EDH presented here. The sheet differs from the conventional vortex sheet by allowing mass and consequently momentum in the sheet

To demonstrate a boundary layer, the Blasius boundary layer is considered which is a special case of flat plate boundary layer. Here, the boundary layer is formed on a semi-infinite flat plate, with a constant free-stream flow. This requires the following assumptions to be made:
\begin{enumerate}
\item Steady state, no time rate of change of hypersurface quantities.
\item No source of hypersurface mass or momentum.
\item No hypersurface body force.
\item Location of the bounds of the boundary layer (or width) does not change with time. 
\item Velocity on the side of EDH next to the stationary wall is zero, ${\bf u}\big|_{n_1}=0$, because of no-slip and no-penetration condition. 
\item The fluid outside the boundary layer is assumed to be irrotational. Therefore, jump in shear stress $ \left[\!\left[{T}^m_{sn} \right]\!\right] = 0 - \tau_w$, where $\tau_w$ is the shear at the wall.
\item Since we are considering a periodic or 2D problem, there is no surface shear $T_{sn}^{(m-1)'}=0$.
\item There is no surface dilatation, $\nabla^{m-1} \bm\cdot u^{(m-1)'}=0$.
 \item Also there is no surface gradient in surface pressure.
\end{enumerate}



By considering these assumptions the {\bf actual extended dividing hypersurface} reduces to:

{\it Mass conservation}
\begin{align}
\nabla_{\hat{\bm s}}^{m-1} \bm \cdot (\rho{\bf u})^{(m-1)'} ~+~ (\rho{\bf u})^m_{n_2} \bm\cdot \left(\nabla_{\hat{\bm s}}^{m-1} (n_2) \right)~ +  ~  (\rho{\bf u})^m \bm\cdot \hat{\bf n} \Big|_{n_2} =  0 
\end{align}

 {\it Momentum conservation}
\begin{align}
\nabla_{\hat{\bm s}}^{m-1} \bm\cdot \left(\rho {\bf u} \otimes {\bf u}\right)^{(m-1)'} 
 + \left(\rho {\bf u} \otimes {\bf u} \right)^m \bm\cdot \hat{\bf n} \Big|^{n_2}       - \left({\bf T}^m \bm\cdot \hat{\bf n}\Big|^{n_2} - {\bf T}^m \bm\cdot \hat{\bf n}\Big|_{n_1}\right)\\ \notag
 \left(\rho {\bf u} \otimes {\bf u}\right)^{m}_{n_2} \bm \cdot \left( \nabla_{\hat{\bm s}}^{m-1} (n_2) \right)  
 -   \left({\bf T}\right)^{m}_{n_2} \bm\cdot \left(\nabla_{\hat{\bm s}}^{m-1} (n_2) \right)  = 0.
\end{align}


Similarly, choosing the location of hypersurface to be at the wall, the {\it Mass conservation} of {\bf effective extended dividing hypersurface} reduces to
, 
\begin{align}
\nabla_{\hat{\bm s}}^{m-1} \bm \cdot (\rho{\bf u})^{(m-1)}  ~ +  ~  (\rho{\bf u})^m \bm\cdot \hat{\bf n} \Big|_{n_0} =  0 
\end{align}
 for a vortex sheet fixed to a stationary wall, since $n_0$ is not varying with space. Here, one thing to note is that $(\rho{\bf u})^{(m-1)} \neq \int \left(\rho{\bf u}\right) dn$ but rather it accounts for the bulk fluid added in the void region $(\int \left(\rho{\bf u}\right)_B dn)$. Hence, $(\rho{\bf u})^{(m-1)} = \int \left(\rho{\bf u}\right) dn -  \int \left(\rho{\bf u}\right)_B dn$. 
 
%

{\it momentum conservation}
\begin{align}
\nabla_{\hat{\bm s}}^{m-1} \bm\cdot \left(\rho {\bf u} \otimes {\bf u}\right)^{(m-1)} 
 + \left(\rho {\bf u} \otimes {\bf u} \right)^m \bm\cdot \hat{\bf n} \Big|_{n_0}       - \left[\!\left[{\bf T}^m \bm\cdot \hat{\bf n}\right]\!\right]_{n_0}  = 0.
\end{align}

Decomposing into normal and tangential components.
\begin{subequations}
\begin{align}
\frac{\partial}{\partial s} \left(\rho u_s u_n\right)^{(m-1)} 
 + \left(\rho u_n^2  \right)^m  \Big|_{n_0}       -  \left[\!\left[p\right]\!\right]_{n_0}  = 0.
\end{align}
\begin{align}
\frac{\partial}{\partial s} \left(\rho  u_s^2\right)^{(m-1)} 
 + \left(\rho u_n u_s  \right)^m \Big|_{n_0}       +\tau_w
 = 0.
\end{align}
\end{subequations}
Assuming flow to be incompressible, ${\bf T}^m \bm\cdot \hat{\bf n} = p{\bf I}$. The mass and momentum conservation equation presented here for an effective EDH, is same as the equation for the vortex-entrainment sheet presented by \cite{Mohseni:19a}. Thereby, we see that the vortex-entrainment sheet is just a special case of the EDH.

The region of validity of the vortex-entrainment sheet was, $Re_x > 100$. From the equations for the EDH, it can be seen that the region of validity of EDH can be extended all the way till the leading edge, if appropriate assumptions are relaxed. As for the leading edge, in order to capture that we would need to consider an extended dividing hypersurface in 1D space. That is the EDH equations describing the leading edge will need to be computed by collapsing the dimension twice in the directions normal to the line of discontinuity.



\subsection{Shock tube problem}
In all the test cases considered so far, the extended dividing hypersurface had no normal velocity, ${\bf u} \bm\cdot \hat{\bf n} = 0$. In the case of the shock tube problem, the shock wave falls under the classification of a physical front. Hence, the shock wave can be represented by an EDH, which propagates in a direction normal to it. 

As done before, we first list all the assumptions
\begin{enumerate}
\item Steady state, no time rate of change of hypersurface quantities.
\item 1D flow, no spatial gradients along the length of the EDH.
\item No source of hypersurface mass and momentum.
\item No hypersurface body force.
\item Thickness of the shock front does not change with time.
\end{enumerate}

By considering these assumptions the {\bf actual extended dividing hypersurface} reduces to:

{\it Mass conservation}

\begin{align}
- \left[\!\left[ (\rho{\bf v})^m \bm\cdot \hat{\bf n} \right]\!\right]_{n_1}^{n_2} ~ + ~ 
  \left[\!\left[ (\rho{\bf u})^m \bm\cdot \hat{\bf n} \right]\!\right]_{n_1}^{n_2} =  0 
\end{align}

{\it Momentum conservation}
 \begin{align}
\left[\!\left[ \left(\rho {\bf u} \otimes {\bf u} \right)^m \bm\cdot \hat{\bf n} \right]\!\right]_{n_1}^{n_2}   - {\left[\!\left[{\bf T}^m \bm\cdot \hat{\bf n}\right]\!\right]_{n_1}^{n_2} } 
+\left[\!\left[ \left(\rho {\bf u}\right)^{m}\frac{d h }{dt} \right]\!\right]_{n_1}^{n_2} = 0,
\end{align}

The mass conservation equation reduces to the Rankine-Hugoniot condition.
\begin{align}
  \left[\!\left[ (\rho ({\bf u} - {\bf v}))^m \bm\cdot \hat{\bf n} \right]\!\right]_{n_1}^{n_2} =  0 
\end{align}

\begin{figure}
\begin{minipage}{1.0\linewidth}
\centerline{
 \includegraphics[width=0.65\textwidth]{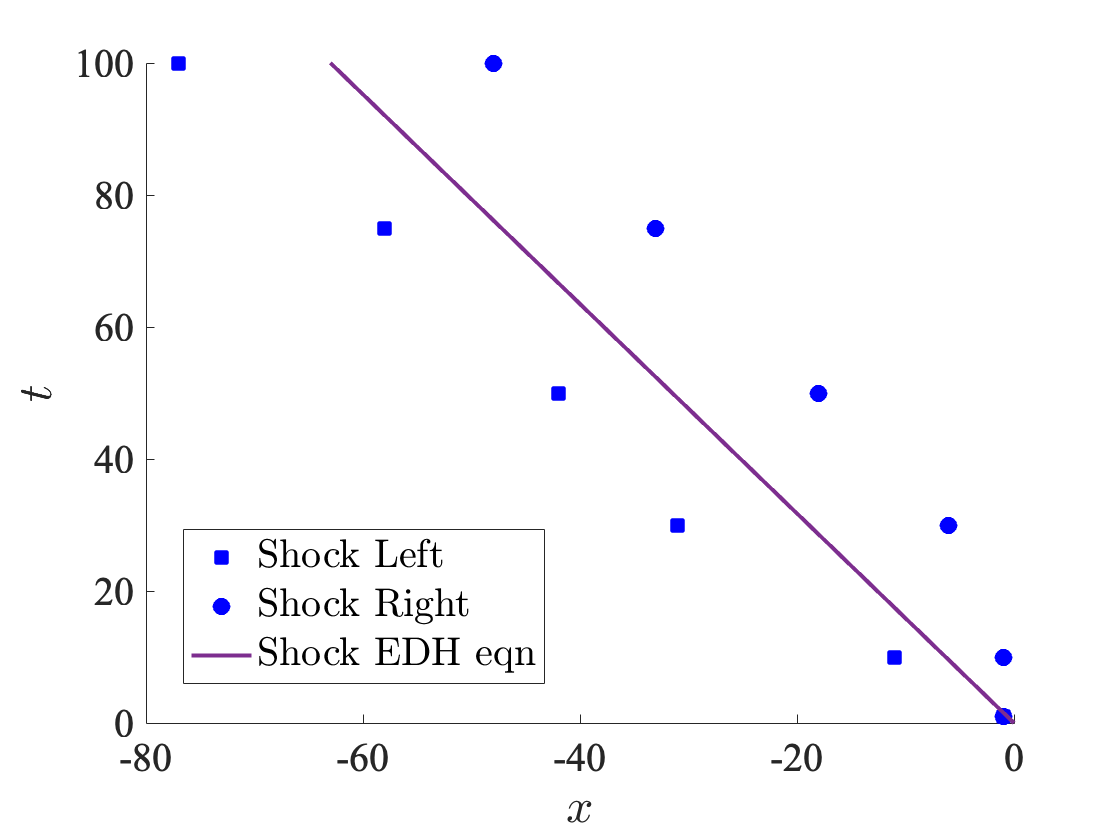}
 }
 \end{minipage}
 \\
 \caption{Comparing displacement vs t, from MD and from Surface mass conservation equations.}
\label{fig:ShockExapn_d_vs_t}
\end{figure}

Since, the two edges of the shock wave are considered to move with the same speed, after reaching steady state. The equation can be rewritten as
\begin{align}
- \left[\!\left[ (\rho)^m \right]\!\right]_{n_1}^{n_2} ({\bf v} \bm\cdot \hat{\bf n}) ~ + ~ 
  \left[\!\left[ (\rho{\bf u})^m \bm\cdot \hat{\bf n} \right]\!\right]_{n_1}^{n_2} =  0. 
\end{align}
Hence, if we know the bulk density and velocity, on the two sides of the shock front, we can then evaluate the speed of the shock wave (${\bf v}\bm\cdot \hat{\bf n}$). Figure \ref{fig:ShockExapn_d_vs_t}, compares the result for the displacement of shock front with time obtained from MD to that obtained using the above EDH equation. The displacement of shock front obtained from EDH equation agrees well with the MD data. Preliminary work related to representing a shock wave as an EDH was done by \cite{Thalakkottor:23a}.

Although the validity of EDH to represent a shock front is being shown for the canonical normal shock, it has the ability to accommodate more general problems. 

\subsection{Collapsing bubble}
This example investigates the evolution of a bubble under an unstable condition of high pressure at infinity. Specifically, it examines the radial evolution of an unstable bubble in an unbounded liquid, which is described by the Rayleigh-Plesset equation \cite{PlessetMS:49a, ProsperettiA:77a, BrennerM:02a, BrennenCE:14a, LealLG:10a}. The evolution of a collapsing bubble is extensively studied in the field of cavitation and multiphase flows. All preceding examples have focused on examining planar interfaces in a steady-state condition. However, this particular example investigates a curved interface within an unsteady state.

The Rayleigh-Plesset (RP) equation is employed to model the time-dependent behaviour of the bubble. However, in typical multiphase flow formulations, the RP equation only considers surface tension as the interfacial property and incorporates the corresponding pressure jump. Consequently, it neglects the interface mass and associated dynamics. In this study, we demonstrate that by incorporating the conservation equation for interfacial (front) mass in conjunction with the Rayleigh-Plesset equation, the evolution of the interface undergoes significant changes.

The Rayleigh-Plesset equation is written as:
\begin{align}
{p_b-p_{\infty}}{} =  \frac{2\gamma}{R} + \rho_l R\frac{d^2 R}{dt^2} + \rho_l \frac{3}{2}\left(\frac{d R}{d t}\right)^2 + 4\mu_l\frac{1}{R}\frac{d R}{d t}
\end{align}
Here, $p_b$ is pressure within the bubble, $p_{\infty}$ pressure in the liquid at infinity, $\rho_l$ density of liquid, $\mu_l$ dynamic viscosity of the liquid, and $R$ is the radius of the bubble or location of the hypersurface (interface). This can be re-written in terms of velocity of the interface, where $u=dR/dt$.
\begin{align}
{p_b-p_{\infty}}{} =  \frac{2\gamma}{R} + \rho_l R\frac{d u}{dt} + \rho_l \frac{3}{2}\left(u\right)^2 + 4\mu_l\frac{1}{R}u
\end{align}

Next, we consider that the interface has a finite mass. The interface mass conservation equation for a system without any mass flux across the interface reduces to:
\begin{equation}
d_t \left( M \right) = \int_S d_t \left (\rho^{m-1}\right) dS = 0
\end{equation}

\begin{figure}
\center
\begin{minipage}{0.45\linewidth}
\centerline{
 \includegraphics[width=1.0\textwidth]{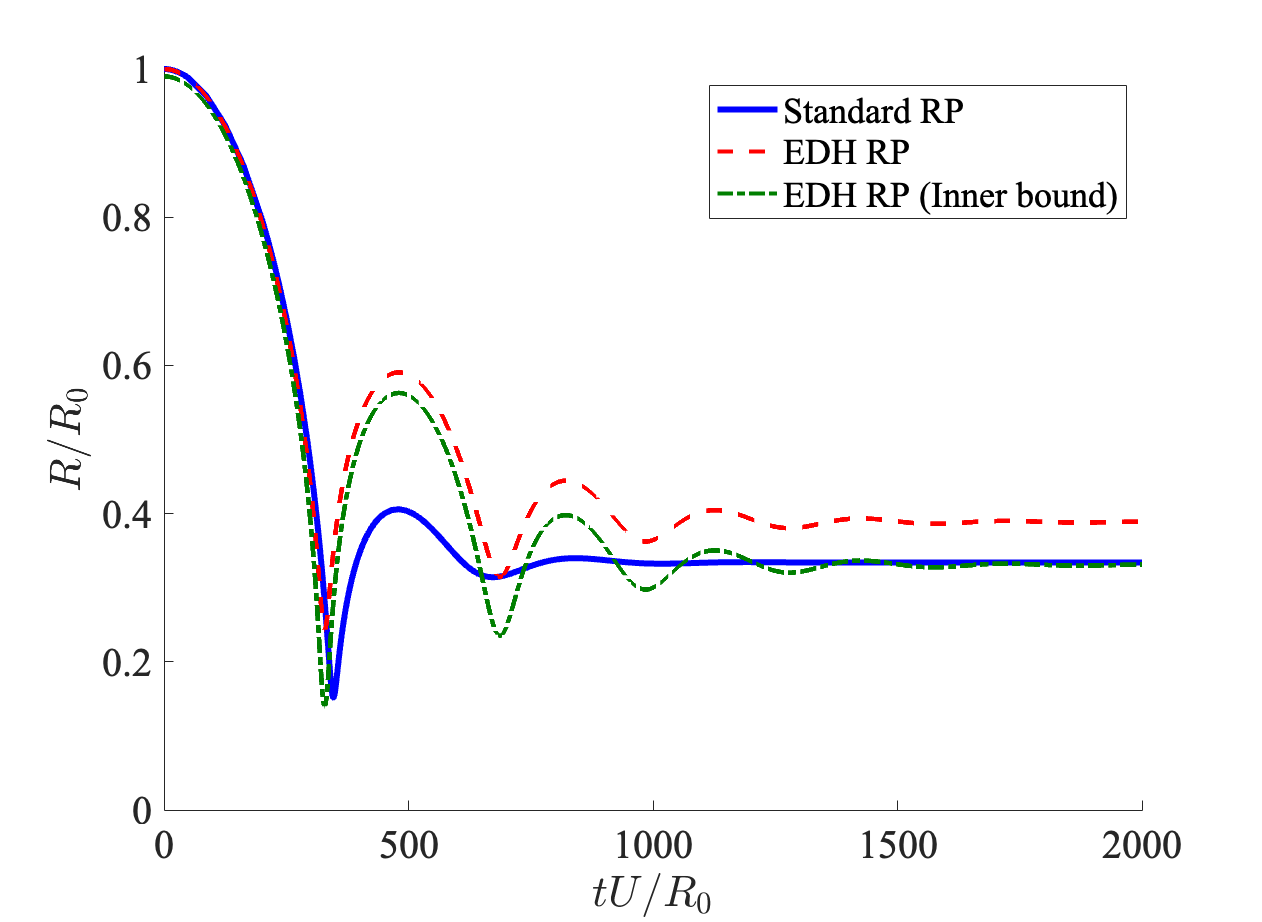}
 }
 \end{minipage}
 \begin{minipage}{0.45\linewidth}
\centerline{
 \includegraphics[width=1.0\textwidth]{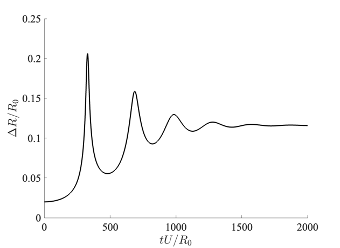}
 }
 \end{minipage}
 \\
 \caption{Comparing the evolution of bubble obtained from EDH RP and standard RP. Evolution of (a) bubble radius and (b) interface thickness.}
\label{fig:BubbCollap_R_vs_t}
\end{figure}

The evolution of the bubble is shown in figure \ref{fig:BubbCollap_R_vs_t}. Comparing the standard RP to EDH RP, which accounts for interface mass and thickness, distinct differences between the evolution of the bubble radius for these two cases are observed. While the rate of the first collapse looks identical between the two models, the first major distinction is in the minimum radius attained. The minimum radius of modified RP equation is larger than that of the RP equation, leading to an improved agreement with experimental results \citep{HwangfuJJ:10a,ColeRH:48a}. This increase is directly a result of the interface having a finite mass and thickness. In order to better understand the cause of this difference it is best to look at fig \ref{fig:BubbCollap_R_vs_t}(a) in conjunction with \ref{fig:BubbCollap_R_vs_t}(b). As the bubble radius decreases, mass conservation dictates that the interface thickness must increase in order to compensate for the reduction in the surface area of the bubble. In other words, geometric expansion dictates the increase in interface thickness, as there is no mass flux into the interface. This means that when the EDH (interface) reaches its minimum radius (red dashed line), the corresponding location of the inner bounds of the interface (green dashed line), would have, in fact, reached the same location as that of a standard RP equation. So, in both the standard and modified RP models, the maximum pressure attained inside the bubble is the same, but since the interface thickness is not the same the bubble radius ends up being different. 

It must be noted that the interface thickness is analytically computed from the initial condition of initial interface thickness and the density sequence (functionality). Hence, in the case of a numerical simulation various fluid and flow properties do no need to be resolved across the interface thickness.


If, in addition to mass conservation, we were to include the momentum conservation of the interface as well, the new RP equation would be.  
\begin{align}
p_b - p_{\infty} = \frac{2\gamma}{R_0} + 4\mu\frac{u_{(r=R_2)}}{R_2^3} +  R_2\frac{\partial u_{(r=R_2)}}{\partial t} - \frac{\rho}{2} u_{(r=R_2)}^2\\ \notag
\frac{\partial (\rho u)^s}{\partial t} + \frac{2(\rho u u)^s}{R_0}
\end{align}
This is outside of the scope of this paper and will be explored in detail in a future work.





\section{Conclusion}
In this paper a systematic derivation of Gibbs' dividing surface from the 3D bulk conservation equations that accurately describe this diffused region is proposed. This generalized dividing surface is referred to as the extended dividing hypersurface (EDH). The EDH equations are derived by collapsing the dimension across the width of the diffused region, mathematically achieved through integration along its width. This mathematical treatment ensures that the EDH is kinematically and dynamically equivalent to the diffused region and represents the real physical front in its entirety.

The selected example problems in this study serve four main purposes within the context of the extended dividing hypersurface (EDH):
A set of canonical example problems are used to validate the EDH model. These example help deduce the following conclusion.

Firstly, they demonstrate that the EDH can accurately capture the dynamics of not only phase or material interfaces, but also other types of fluid and flow fronts. This highlights the versatility of the EDH in representing a wide range of dynamic phenomena.

Secondly, the problems emphasize that the distribution of monotonicity within a front, as commonly described in existing literature, is just one of several possible functionalities that the EDH can effectively represent. This implies that the EDH can offer alternative representations of fluid fronts that may differ from the conventional understanding of monotonicity distribution.

Thirdly, the selected problems showcase the relationship between the flux of $m$- dimensional quantities and the ($m$-1)-dimensional quantities, known as hypersurface quantities. This highlights how the coupling between these quantities can lead to hypersurface dilatation, even in incompressible hypersurface flows. 
Thereby revealing a counterintuitive aspect of hypersurface dynamics.

Finally, these problems underscore the significance of acknowledging the mass of the front and its impact on the dynamics of the front and the adjacent bulk media. By capturing the dynamics of the front, the EDH model provides a comprehensive understanding of the system under consideration.

Overall, this study lays down the framework of extending Gibbs' dividing surface by systematically deriving the governing equations. This allows the extension of the dividing surface concept, referred to here as extended dividing hypersurface (EDH).
It demonstrates the capability of the extended dividing hypersurface (EDH) to accurately represent various fluid and flow fronts beyond traditional interfaces. Through the examination of canonical problems, the authors validate the EDH model and its generalization, contributing valuable insights into its ability to capture the dynamics of the front and the surrounding bulk fluid.

\section{Acknowledgment}
This research was partially supported by the National Science Foundation and Office of Naval Research. 

\bibliographystyle{jfm}

\bibliography{RefA2.bib}

\end{document}